\documentclass[usenatbib,useAMS english]{mnras}
\usepackage[fleqn]{amsmath}
\usepackage{array,multirow,graphicx}
\usepackage{txfonts}
\usepackage{lmodern}
\bibpunct{(}{)}{;}{a}{}{,}
\usepackage{floatrow}
\usepackage{array}
\usepackage{relsize}
\usepackage{bm}
\usepackage{listings}
\usepackage{url}
\usepackage{upgreek}
\usepackage[font=small,labelfont=bf]{caption}

\graphicspath{{./figs/}}
\newcommand{\zo}{$z\!=\!0$}
\newcommand{\HI}{H\,{\sc i}} % HI with correct lettering
\newcommand{\mHI}{$m_{\rm H\,{\LARGE{\textsc i}}}$} % HI mass
\newcommand{\Htwo}{H$_{2}$}
\newcommand{\mHtwo}{$m_{\rm H_2}$}
\newcommand{\ds}{{\sc Dark Sage}}

\newfloatcommand{capbtabbox}{table}[][\FBwidth]

\def\app#1#2{%
\mathrel{%
\setbox0=\hbox{$#1\sim$}%
\setbox2=\hbox{%
\rlap{\hbox{$#1\propto$}}%
\lower1.1\ht0\box0%
}%
\raise0.25\ht2\box2%
}%
}

\title[H$_2$ \& environment of IllustrisTNG galaxies]%
{Molecular hydrogen in IllustrisTNG galaxies: carefully comparing signatures of environment with local CO \& SFR data}
\author[A.~R.~H.~Stevens et al.]{Adam R.~H.~Stevens,$^{1,2}$\thanks{E-mail: adam.stevens@uwa.edu.au}
Claudia del P.~Lagos,$^{1,2}$ Luca Cortese,$^{1,2}$ Barbara Catinella,$^{1,2}$
\newauthor  Benedikt Diemer,$^3$ Dylan Nelson,$^{4,5}$ Annalisa Pillepich,$^6$ Lars Hernquist,$^7$ Federico Marinacci$^8$
\newauthor  and Mark Vogelsberger$^9$ 
\\
$^1$International Centre for Radio Astronomy Research, The University of Western Australia, Crawley, WA 6009, Australia\\
$^2$Australian Research Council Centre of Excellence for All Sky Astrophysics in 3 Dimensions (ASTRO 3D)\\
$^3$Department of Astronomy, University of Maryland, College Park, MD 20742, USA\\
$^4$Max-Planck-Institut f\"{u}r Astrophysik, 85741 Garching, Bayern, Germany\\
$^5$Institut f\"{u}r theoretische Astrophysik, Zentrum f\"{u}r Astronomie, Universit\"{a}t Heidelberg, 69120 Heidelberg, Baden-W\"{u}rttemburg, Germany\\
$^6$Max-Planck-Institut f\"{u}r Astronomie, 69117 Heidelberg, Baden-W\"{u}rttemburg, Germany\\
$^7$Institute for Theory and Computation, Harvard-Smithsonian Center for Astrophysics, Cambridge, MA 02138, USA\\
$^8$Department of Physics \& Astronomy, University of Bologna, 40129 Bologna, Italy\\
$^9$Department of Physics, Massachusetts Institute of Technology, Cambridge, MA 02139, USA}

\begin{document}

\defcitealias{gk11}{GK11}
\defcitealias{gd14}{GD14}
\defcitealias{k13}{K13}
\defcitealias{stevens19}{S19}

\pagerange{\pageref{firstpage}--\pageref{lastpage}} \pubyear{2021}

\maketitle

\label{firstpage}

\begin{abstract}
We examine how the post-processed content of molecular hydrogen (\Htwo) in galaxies from the TNG100 cosmological, hydrodynamic simulation changes with environment at \zo, assessing central/satellite status and host halo mass.   We make close comparisons with the carbon monoxide (CO) emission survey xCOLD GASS where possible, having mock-observed TNG100 galaxies to match the survey's specifications.  For a representative sample of host haloes across $10^{11}\!\lesssim\!M_{\rm 200c}/{\rm M}_{\odot}\!<\!10^{14.6}$, TNG100 predicts that satellites with $m_*\!\geq\!10^9\,{\rm M}_{\odot}$ should have a median deficit in their \Htwo~fractions of $\sim$0.6\,dex relative to centrals of the same stellar mass. Once observational and group-finding uncertainties are accounted for, the signature of this deficit decreases to $\sim$0.2\,dex.
Remarkably, we calculate a deficit in xCOLD GASS satellites' \Htwo~content relative to centrals of 0.2--0.3\,dex, in line with our prediction.
We further show that TNG100 and SDSS data exhibit continuous declines in the average star formation rates of galaxies at fixed stellar mass in denser environments, in quantitative agreement with each other.
By tracking satellites from their moment of infall in TNG100, we directly show that atomic hydrogen (\HI) is depleted at fractionally higher rates than \Htwo~on average.
Supporting this picture, we find that the \Htwo/\HI~mass ratios of satellites are elevated relative to centrals in xCOLD GASS.
We provide additional predictions for the effect of environment on \Htwo\,---\,both absolute and relative to \HI\,---\,that can be tested with spectral stacking in future CO surveys.
\end{abstract}


% ============================================================ %
% ============================================================ %
% ============================================================ %
% ============================================================ %
% ============================================================ %
% ============================================================ %
% ============================================================ %

\section{Motivation and Background}
\label{sec:intro}

Complementing pioneering works that date back nearly half a century \citep*[see reviews by][]{haynes84,blanton09}, there has been a flurry of studies in the last decade in particular that highlight how galaxies' cold gas can be affected by their environment, be it in terms of their integrated content with statistical samples from observations \citep{chung09,cortese11,catinella13,boselli14c,jaffe16,stark16,brown17,janowiecki17} or simulations/models \citep{tecce10,marasco16,sb17,cora18,stevens18,stevens19}, or in resolved detail with small numbers \citep{moretti18,boselli19,ram19,jachym19,yun19}.  The vast majority of these works have focussed on the atomic hydrogen (\HI) in galaxies\,---\,the bulk of a galaxy's cold gas\,---\,with a generalized consensus that local galaxy density anti-correlates with galaxy \HI~content.  While some attention has also been given to molecular gas, studies of its relation with galaxy environment are less advanced.  This is understandable, given that molecular emission is more challenging to measure at low redshift.  While \HI~is readily studied through its ubiquitous radio emission at a wavelength of $\sim$21\,cm, \Htwo~(molecular hydrogen) lacks a permanent dipole, and is notoriously difficult to detect from direct electromagnetic emission or absorption.  This means relying on emission of other molecules thought to trace \Htwo: in particular, carbon monoxide \citep*[CO\,---\,see the review by][]{bolatto13}.

To understand how galaxy environment affects any galaxy property, it is vital to first distinguish between centrals and satellites.  Central galaxies reside in the global minimum of a halo's potential well (i.e.~at the halo's \emph{centre}) and are the most massive galaxy in that halo.  In practice, they are often identified observationally as the galaxy with the greatest stellar mass or luminosity in a linked group \citep[various algorithms exist for doing the linking\,---\,e.g.][]{yang07,robo11,tempel16,tinker20}.  Satellite galaxies make up the remaining, subordinate population of galaxies in haloes.  They were once centrals themselves, but have since fallen into haloes greater than their own, per the hierarchical assembly of structure in the Universe under the $\Lambda$CDM cosmological model \citep[c.f.][]{white78}.  Centrals can have any number of associated satellites (including zero).  A central and its environment evolve in tandem, while satellites suffer environmental effects that centrals are scarcely affected by.  These effects include, but are not limited to, ram-pressure stripping, starvation/strangulation of fresh gas, and tidal stripping \citep*[see][respectively]{gunn72,larson80,moore99}.

Environmental processes that strip the interstellar media of galaxies are expected to impact the atomic-gas content of those galaxies more than their molecular gas.  This is because the outskirts of galaxies are more susceptible to tides and ram pressure, where the atomic-to-molecular ratio is almost always much higher \citep{leroy08,bigiel12,sb17}.  Nevertheless, once stripping becomes strong enough, \Htwo~in satellites is expected to be lost as well.  

While the number of works targeting variation in molecular-gas content with environment pale in comparison to those of atomic gas, there were efforts to study the effect of cluster environments on galaxies' molecular gas at low redshift as far back as $\sim$30 years ago \citep[e.g.][]{kenney89,casoli91}.
Early conclusions about whether these galaxies are \Htwo-poor or not was later brought into question, in part due to selection effects and uncertainty surrounding the CO-to-\Htwo~conversion factor \citep[see][and references therein]{boselli14c}.
Debate has since ensued as to whether dense environments typically produce \Htwo-poor galaxies.
At \zo, based on the fact that the most `\HI~deficient' galaxies in the \emph{Herschel} Reference Survey also exhibit some deficiency in \Htwo, \citet{boselli14c} argue that cluster satellites are at least partially stripped of \Htwo.  
Evidence for direct, resolved stripping of CO has also been observed in a handful of galaxies \citep{lee18,moretti18,cramer19,jachym19}, while additional cases of molecular disturbances have been found in Virgo \citep{lee17}.
Broadly speaking, at intermediate and higher redshifts, reports are even more varied, with some suggestions that (proto)clusters favour \Htwo-poor galaxies \citep{jablonka13,castignani18,coogan18}, others that they favour \Htwo-rich galaxies \citep{noble17,hayashi18}, others that they favour \Htwo-normal galaxies \citep{dannerbauer17,lee17b,rudnick17,darvish18,wu18}, or a mass-dependent/phase-space-dependent combination thereof \citep{hayashi17,wang18,tadaki19}.  

Part of the challenge of drawing empirical conclusions about the interplay between galaxy environment and \Htwo~content comes from data in the literature being heterogeneous and/or limited in sample size.  
To make robust theoretical predictions for an environment--\Htwo~relation is also non-trivial, as it ideally involves cosmological, hydrodynamic simulations of significant volume (box lengths of $\gtrsim\!100$\,Mpc), sufficient resolution (baryonic mass elements of $\lesssim\!10^6\,{\rm M}_{\odot}$), and an accounting of the multi-phase nature of galactic gas that differentiates between its ionized, atomic, and molecular components.
In recent years especially, there has been significant development in the gas-phase decomposition of hydrodynamic simulations, leading to a string of model predictions for the atomic- and molecular-gas content of galaxies \citep{dave13,dave20,tomassetti14,tomassetti15,lagos15b,rafie15,rafie19,rahmati15,bahe16,marasco16,crain17,marinacci17,diemer18,diemer19,rodriguez19,stevens19,stevens19b,schabe20}.  All these simulations offer laboratories of a sort that allow us to systematically study the causal impact between a wide range of galaxy properties.  
To that point, this paper is focussed on the relationship between the environment and \Htwo~content of galaxies,
making use of the TNG100 simulation of the IllustrisTNG\footnote{Illustris: The Next Generation\,---\,\url{http://www.tng-project.org/}} suite.  Where possible, we compare and contrast our simulation results with data from observational molecular gas studies, most notably the xCOLD GASS program \citep{saintonge17}, and otherwise make predictions that future surveys may be able to test.

This paper serves as a sequel to \citet[][hereafter S19]{stevens19}.  Where \citetalias{stevens19} targeted the effect of environment on the \HI~properties of galaxies, here we focus equivalently on \Htwo, using the same simulation (TNG100) and an overlapping observational galaxy sample (xCOLD GASS).  We describe both the simulation and data in Section \ref{sec:method}, including our method for `mock observing' the simulated galaxies consistently with the survey specifications.  In Section \ref{sec:comparison}, we compare the overall \Htwo~properties of galaxies from TNG100 and xCOLD GASS at \zo~as a function of stellar mass, serving as a reference point for our main results, which are subsequently presented in Section \ref{sec:env}.  There, we slice the data by environment in several ways, not only presenting changes to \Htwo~with these slices, but also how changes to \HI~and star formation scale with them.
We discuss several caveats in Section \ref{sec:elephants}, and offer final remarks and conclusions in Section \ref{sec:conc}.
Ancillary information regarding our methods and results is provided in Appendices \ref{app:anc} \& \ref{app:app}.

% ============================================================ %
% ============================================================ %
% ============================================================ %
% ============================================================ %
% ============================================================ %
% ============================================================ %
% ============================================================ %

\section{Data and methods}
\label{sec:method}

\subsection{The TNG100 simulation}
\label{ssec:tng}

The focus of this paper is on the results of the TNG100 simulation \citep{pillepich18b,nelson18,nelson19,marinacci18,naiman18,springel18}, part of the IllustrisTNG magnetohydrodynamic, cosmological simulation suite.  The simulation was run with the {\sc arepo} code \citep{springel10}, and follows the standard $\Lambda$CDM cosmological model, with parameters based on the \citet{planck16} results: $\Omega_m \! = \! 0.3089$, $\Omega_{\Lambda} \! = \! 0.6911$, $\Omega_b \! = \! 0.0486$, $h \! = \! 0.6774$, $\sigma_8 \! = \! 0.8159$, $n_s \! = \! 0.9667$.  The TNG model includes a range of sub-grid prescriptions that are designed to accommodate relevant astrophysical processes in the formation and evolution of galaxies: for example, gas cooling, star formation, massive black-hole growth, and feedback from both stars and active galactic nuclei \citep{weinberger17,pillepich18a}.  These are based on the earlier Illustris simulation's sub-grid models \citep[see][]{vogelsberger13,vogelsberger14a,vogelsberger14b,genel14,torrey14}.  TNG100 employs a periodic box of length $75\,h^{-1} \! \simeq \! 110\,{\rm cMpc}$, containing $1820^3$ dark-matter particles of mass $7.5 \! \times \! 10^6\,{\rm M}_{\odot}$, and $1820^3$ initial gas cells of typical mass $\sim$$1.4 \! \times \! 10^6\,{\rm M}_{\odot}$.  The minimum gravitational softening scale for gas\,---\,achieved only in galaxy centres\,---\,is 0.19\,kpc.

Gravitationally bound structures in the simulations were found with the {\sc subfind} algorithm \citep{springel01,dolag09}.  These `subhaloes' exist within haloes that are built with a friends-of-friends (FoF) particle group finder.  The most massive subhalo in a FoF group is dubbed the `central', and the remainder are `satellites'.  We follow the nomenclature of \citetalias{stevens19} when describing the integrated properties of galaxies in the simulation.  That is, `inherent' galaxy properties are defined as the subset of particles/cells in a subhalo that also lie within the spherical aperture given by the `baryonic mass profile' criterion (`BaryMP' for short) of \citet{stevens14}.  `Mock' properties depend on the dataset being compared to.  In this paper, these relate to the xCOLD GASS survey and Sloan Digital Sky Survey (SDSS\,---\,\citealt{york00}).  We describe the mocking process for this in Section \ref{ssec:mock}.

All gas cells in the simulation are post-processed in order to calculate their mass fractions in the form of atomic and molecular hydrogen.  The method is described in \citetalias{stevens19}, which builds from \citet{lagos15b} and \citet{diemer18}.  In this paper, we follow the same three prescriptions used by \citetalias{stevens19}, namely those based on the works of \citet[][hereafter GK11]{gk11}, \citet[][hereafter K13]{k13}, and \citet[][hereafter GD14]{gd14}.  All three prescriptions rely principally on the local UV field strength (which we model for each galaxy in three dimensions\,---\,see \citealt{diemer18} for details), local metallicity (a proxy for dust), and local surface density (translated from volumetric gas densities through the Jeans approximation) to calculate the \HI/\Htwo~mass ratio of each gas cell.

For TNG100 results at \zo, the same galaxy sample analysed in \citetalias{stevens19} is analysed in this paper.  That is, only subhaloes with $m_* \! \geq \! 10^9\,{\rm M}_{\odot}$ and dark-matter fractions above 5\% are included.

% ============================================================ %

\subsection{The xCOLD GASS survey}
\label{ssec:xcold}

The xCOLD GASS\footnote{eXtended Carbon monOxide Legacy Database for the \emph{Galaxy evolution explorer} Arecibo `Sloan digital sky survey' Survey} program comprises two observational surveys of the CO(1--0) line emission in galaxies with the IRAM\footnote{Institute for Radio Astronomy in the Millimetre range} 30-m telescope.  The surveys were designed to measure (or place upper limits on) the H$_2$ content of a representative sample of galaxies over $>$2\,dex in stellar mass.  The original COLD GASS survey \citep{saintonge11} targeted 350 galaxies with $m_* \! \geq \! 10^{10}\,{\rm M}_{\odot}$ in the redshift range $0.025 \! \leq \! z \! \leq \! 0.05$.  These galaxies were a subset of those observed in 21-cm \HI~emission in the GASS survey \citep{catinella10}.  Each galaxy was observed sufficiently long to either detect the CO(1--0) line or place an upper limit of $\sim$$0.015\,m_*$ on $m_{\rm H_2}$.  The second survey\,---\,COLD GASS-low \citep{saintonge17}\,---\,observed galaxies with $10^{9} \! < \! m_*/{\rm M}_{\odot} \! < \! 10^{10}$ at $0.01 \! \leq \! z \! \leq \! 0.02$, with detections in CO for $m_{\rm H_2} \! \gtrsim \! 0.025\,m_*$.  Most galaxies in COLD GASS-low, but not all, also have \HI~data from GASS-low \citep{catinella18}.  In total, xCOLD GASS comprises 532 galaxies, 477 of which are also in xGASS (GASS + GASS-low).  As per \citet{catinella18}, we refer to the subsample of galaxies in both surveys as `xGASS-CO'.  All \Htwo~masses from xCOLD GASS shown in this work assume the CO-to-\Htwo~conversion factor of \citet{accurso17}, which depends on both the metallicity and star formation rate (SFR) of a galaxy.

All xCOLD GASS data were sourced from the publicly available catalogue.  We note that tags for galaxies being centrals and satellites are not included in this dataset, but they are for xGASS (see Section \ref{ssec:sdss}).  Where we present centrals and satellites for xCOLD GASS, we therefore use the xGASS-CO subset.
We also note that \Htwo~masses given in the public dataset and presented in \citet{saintonge17} are actually $1.36\,m_{\rm H_2} \! \simeq \! m_{\rm H_2} \! + \! m_{\rm He}$ \citep[see section 1 of][]{accurso17}.  In this work, $m_{\rm H_2}$ means the mass of molecular hydrogen \emph{only}. 
We therefore subtract the helium contribution from all xCOLD GASS measurements.

Measured fluxes must undergo `beam corrections' (often also referred to as `aperture corrections' in the literature) to estimate the intrinsic flux of an object.  The beam corrections assume that (i) the face-on, two-dimensional \Htwo~half-mass radius, $r_{\rm H_2}^{\rm half}$, and half-SFR radius, $r_{\rm SFR}^{\rm half}$, are equivalent for each galaxy (where the latter is measured independently); (ii) the \Htwo~surface density of every galaxy falls exponentially with radius; and (iii) the IRAM beam shape is a Gaussian.  
We can analytically write the beam correction factor as
\begin{subequations}
\label{eq:corr}
\begin{multline}
f^{\rm beam}_{\rm corr} = \left[ 2\uppi\, r_d^2\, \cos(i) \right]^{-1}  ~\times \\ 
\int^{\infty}_{-\infty}\int^{\infty}_{-\infty} \exp\left( -\frac{x^2 + y^2}{2\sigma_{\rm beam}^2} - \frac{\sqrt{x^2 + y^2\sec^2(i)}}{r_d} \right) {\rm d}x\, {\rm d}y\,,
\end{multline}
\begin{equation}
r_d  = 0.596\, r_{\rm SFR}^{\rm half}\,,
\end{equation}
\begin{equation}
\sigma_{\rm beam} = {\rm FWHM}\,/\!\sqrt{8\ln(2)}\,,
\end{equation}
\end{subequations}
where $i$ is the galaxy's inclination, and `FWHM' refers to the full width at half maximum of the beam\,---\,often simply referred to as the `beam size'\,---\,where the beam size of IRAM is 22\,arcsec at the frequency of the CO(1--0) observations.
Further details regarding the beam correction method can be found in section 2.2 of \citet{saintonge12}.  
Both the beam-corrected and uncorrected \Htwo~masses are used in this work.
For a given galaxy, we assume the conversion factor from CO luminosity to $m_{\rm H_2}$ is the same for the corrected and uncorrected quantities.

Because xCOLD GASS is not complete in terms of stellar mass (even though it is representative), whenever we calculate statistics such as running percentiles or means, we assign weights to each galaxy to compensate for the relative galaxy count in adjacent 0.1-dex slices of stellar mass.  The expected frequency is based on the ratio of galaxy counts in TNG100 for those same bins.  This method follows \citetalias{stevens19}, which is very similar to that used in \citet{catinella18}.

\begin{figure}
\centering
\includegraphics[width=\textwidth]{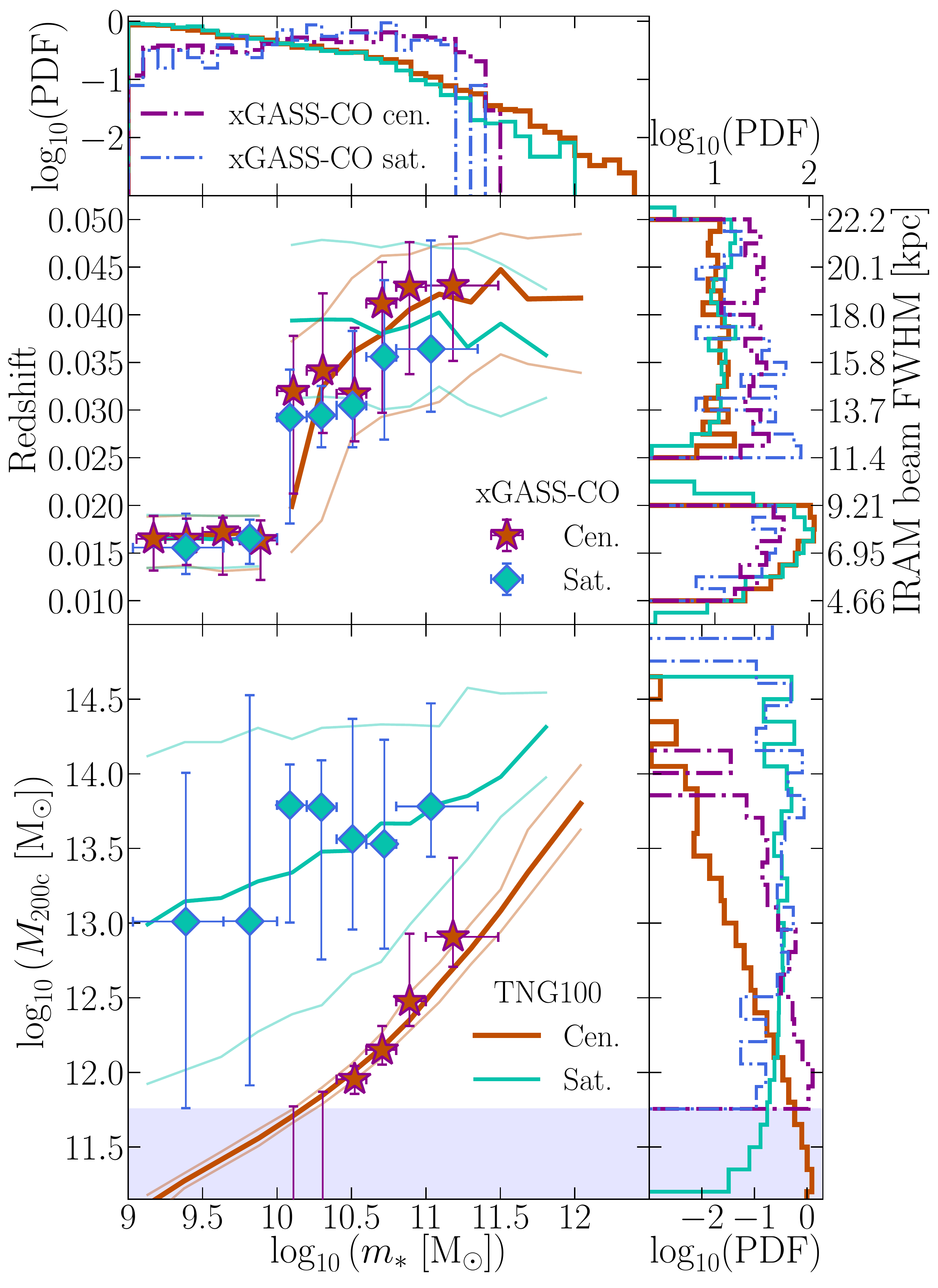}
\caption{Distributions of stellar masses, parent halo masses, and redshifts of galaxies used in this work, split into centrals and satellites.  Points and thick lines in the main panels are respective medians for xGASS-CO and the `mock observed' TNG100 galaxies, per the procedure summarized in Section \ref{ssec:mock} (excluding Section \ref{ssec:contam} in this case). Thin lines and vertical error bars give the corresponding 16$^{\rm th}$ and $84^{\rm th}$ percentiles. Horizontal error bars for xGASS-CO points cover the full bin width.  For each marked redshift, the equivalent physical scale on which \Htwo~masses are measured is given on the right-hand side (FWHM = full width at half-maximum). Smaller panels give the one-dimensional projected probability distribution functions (PDFs) along each axis. The shaded region in the bottom panels is where halo masses are not explicitly computed for xGASS-CO galaxies.  For this reason, points for xGASS-CO centrals drop off at low stellar masses; there do exist galaxies at those masses, but they cannot meaningfully be shown on these axes.}
\label{fig:zdist}
\end{figure}

We note that xCOLD GASS was \emph{not} designed for environmental galaxy studies.  In truth, neither the sample size nor detection threshold are ideal for this use.  Nevertheless, there is a sufficiently representative range of environments covered by the two surveys (see the bottom panels of Fig.~\ref{fig:zdist}), and they presently remain unrivalled in their combined number counts and depth of CO surveys at $z\!\simeq\!0$.  While we stress that we push these data to their usage limit, xCOLD GASS persists as the most relevant dataset in the literature for a point of comparison for this paper.

% ============================================================ %

\subsection{SDSS data}
\label{ssec:sdss}

All stellar masses and SFRs from observational data used in this paper (i.e.~for xCOLD GASS and the additional SDSS sample we describe below) originate from the MPA--JHU%
\footnote{Max Planck institute for Astrophysics--Johns Hopkins University}
catalogue from SDSS Data Release 7 \citep{abazajian09}.  These properties have been adjusted for a \citet{chabrier03} stellar initial mass function (IMF) and \citet{planck16} cosmology as necessary (to be consistent with the TNG model).  Stellar masses are based on fits to each galaxy's spectral energy distribution, as derived from SDSS's broadband photometry (\citealt{salim07}; also see \citealt{kauffmann03}).
SFRs rely on fibre spectroscopy, but have had a correction applied for the small fibre aperture \citep{salim07}.  SFRs for actively star-forming galaxies use H$\alpha$ emission as their primary indicator, which canonically probe historical time-scales of order $10^7$\,yr. For passive galaxies, estimated SFRs are related to the 4000-$\AA$ break \citep{brinchmann04}, which probes an average time-scale of order $10^9$\,yr.
Central/satellite assignments and host halo masses are based on the \citet{yang07} group finder.  For xGASS, this includes tweaks outlined in \citet{janowiecki17}.  Otherwise, we use the `modelB' catalogue.

In Section \ref{sec:sf} and Appendix \ref{app:plane}, we use a subsample of galaxies from SDSS that is much larger than xCOLD GASS to look at environmental effects on galaxies' specific star formation rates (sSFRs).  This sample is volume-limited over the redshift range $0.02 \! \leq \! z \! \leq \! 0.05$, completely covers stellar masses from $10^9$ to $10^{11.5}$\,M$_\odot$, and only includes galaxies with {\tt SFR\_FLAG\_tot} = 0.  This is taken from the sample used in \citet{brown17}, which was also used in \citetalias{stevens19}.

% ============================================================ %

\subsection{Mock observations}
\label{ssec:mock}

\subsubsection{\Htwo~properties}
\label{ssec:H2mock}

A limiting factor in single-dish surveys like xCOLD GASS is that there is a finite beam size that can often be smaller than the galaxies of interest.  And as raised in Section \ref{ssec:xcold}, beam corrections are model-dependent; that is, one must assume how CO (and H$_2$) is distributed in galaxies to estimate the fraction of flux that the beam is insensitive to.  In principle, when comparing to simulations, this assumption can be circumvented by instead applying the same beam form to the simulated galaxies, using their known gas distribution, and comparing with \emph{directly} observed fluxes of equivalent galaxies.  We refer to this strategy as `mock observing'.

Our method for mock-observing galaxies in TNG100 is based on \citetalias{stevens19}; for \Htwo~masses, the process is very similar to that used for calculating mock \HI~masses in \citetalias{stevens19}.  First, we need a means of translating between angular scales and physical scales.  To this end, each FoF group in the \zo~snapshot of TNG100 is given a random redshift that is drawn from the probability distribution function of central galaxies in the xGASS survey at the appropriate stellar mass.  This redshift is directly turned into a cosmological distance.  Distances to satellite galaxies in each FoF group are taken as that of their central plus their inherent $z$-coordinate displacement from their central in the simulation.  Where the central has $m_* \! \geq \! 10^{10}\,{\rm M}_{\odot}$ but has associated satellites with $m_* \! < \! 10^{10}\,{\rm M}_{\odot}$, the host halo is assigned a second redshift, used exclusively for those satellites. This second redshift is drawn from the distribution of \emph{all satellites with} $m_* \! < \! 10^{10}\,{\rm M}_{\odot}$ in the xGASS survey.  This ensures that the hard mass transition from galaxies in the COLD GASS and COLD GASS-low surveys is mimicked in the mock.  
Note the subtle difference here from \citetalias{stevens19}; up to three redshifts were assigned to TNG100 haloes in \citetalias{stevens19} to account for the overlapping mass range in GASS and GASS-low.  The mass overlap in COLD GASS and COLD GASS-low totals less than a handful of galaxies, and is therefore negligible for our considerations.

For our mock IRAM beam, we maintain the assumption that H$_2$ is captured with a 100\% response at the beam's centre, and that the response falls off with projected radius as a Gaussian with a standard deviation of 9.36\,arcsec (FWHM = 22\,arcsec).  
For each galaxy in TNG100, we use its assigned distance to convert this to physical units.  All gas cells in the FoF group are then assigned a weight based on this response and their $xy$ displacement from the galaxy centre (taken as the baryonic centre of potential) to measure the integrated \Htwo~mass of the galaxy of interest (this is done for each galaxy in the FoF group individually).  All gas cells that are outside a threshold line-of-sight (always the same as the $z$-direction) relative velocity to the galaxy centre are then eliminated (equivalent to reassigning their mass weight to zero), as these would not contribute to the theoretical CO line.  This threshold is set by a Tully--Fisher relation that considers inclination (equation 6 of \citetalias{stevens19}; based on \citealt{tully77,reyes11}).

For completeness, we show the two-dimensional redshift--stellar mass plane for the observational data and our TNG100 mock in Fig.~\ref{fig:zdist} (the `mock' properties incorporate Section \ref{ssec:other}).  Consideration of the redshift distribution of observational sources is important, not because we expect significant redshift evolution over the probed range, but because the beam implies significantly different physical `apertures' across the redshift range.  The upper panels of Fig.~\ref{fig:zdist} are directly comparable to those of fig.~2 in \citetalias{stevens19}.  The main differences here are that (i) we are restricted to xGASS-CO rather than the full xGASS sample, (ii) mock redshifts for TNG100 satellites with $m_*/{\rm M}_{\odot}\!\in\!(10^{10},10^{10.23})$ have changed, and (iii) we note the relatively much smaller IRAM beam size on the right-hand side of the $y$-axis.  We caution that about half the galaxies with $m_* \! < \! 10^{10}\,{\rm M}_{\odot}$ have a mock beam of ${\rm FWHM} \! < \! 8\,{\rm kpc}$ applied to them (with an absolute minimum FWHM of 4.66\,kpc).  The potential concern is that we are approaching the scale of the simulation's force resolution; for force calculations involving stellar and dark-matter particles, the softening length is 0.74\,kpc.  This effectively means the distribution of mass within the inner 2\,kpc of a TNG100 galaxy is unreliable \citep{pillepich18b}.

\begin{figure*}
\centering
\includegraphics[width=\textwidth]{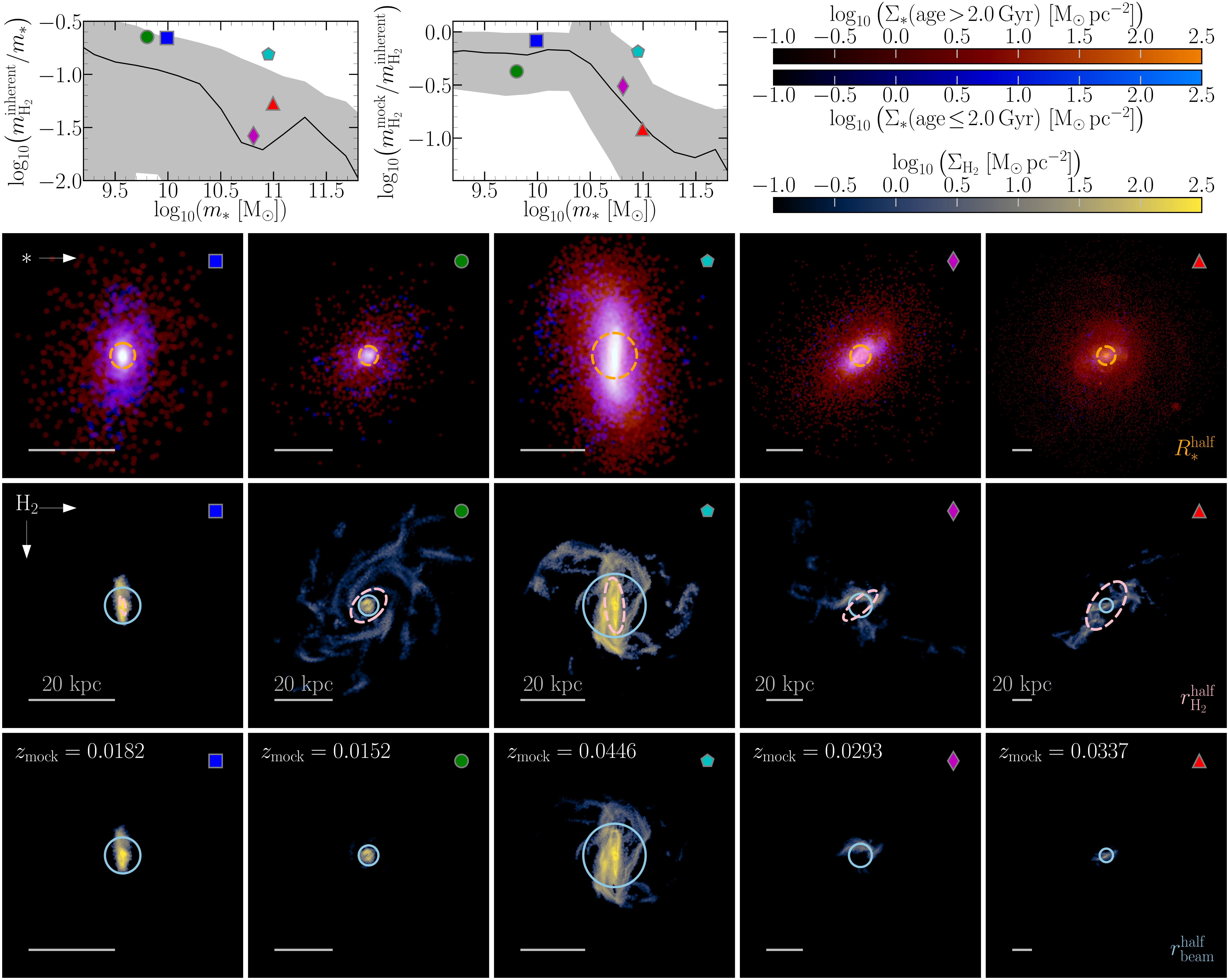}
\caption{Images of five example central galaxies from TNG100.  Two were selected to have stellar masses just under $10^{10}\,{\rm M}_{\odot}$, the other three just under $10^{11}\,{\rm M}_{\odot}$.  All were selected to have an \Htwo~fraction above the median for their stellar mass, as indicated in the top-left panel.  The final selection was to ensure a range of differences between the mock and inherent \Htwo~masses, as seen in the top-centre panel.  In these two panels, the black line is the median for all TNG100 galaxies, while the shaded region covers the 16th--84th percentile range. The top-left panel is directly comparable to the middle panel of Fig.~\ref{fig:H2Frac}.  Each coloured symbol represents a specific galaxy, which can be connected to each column of images.  The first row of images shows the galaxies' stellar content.  Two colours are used here:~red to indicate stellar particles more than 2\,Gyr old, and blue for younger stars.  Dashed, orange circles have radii that match the three-dimensional stellar half-mass radius of each galaxy.  The middle row of images shows all the H$_2$ associated with the galaxy.  Dashed, pink ellipses are the face-on, two-dimensional H$_2$ half-mass radii of the galaxies, each inclined and oriented based on the galaxy's inherent inclination.  The diameter of each solid, sky-blue circle equals the full width at half-maximum for the mock IRAM beam.  The bottom row of images apply the beam response to the H$_2$.  These are therefore not `mock images', but rather a visualization of the mass that contributes to the mock H$_2$ masses.  For visual simplicity, we approximate both stellar particles and gas cells as uniform spheres.  The radius assumed for stellar particles equals the softening scale.  For gas, each cell's recorded volume (mass $\div$ density) is conserved.  
Each frame has dimensions $2R_{\rm BaryMP} \times 2R_{\rm BaryMP}$ (see \citealt{stevens14} and Section \ref{ssec:tng} of this paper).  The \citetalias{gd14} \HI/\Htwo~prescription is used throughout this figure.  From left to right, the subhalo IDs of these galaxies as recorded in the TNG100 database are 482614, 465842, 424255, 408935, and 308060.}
\label{fig:images}
\end{figure*}

To help visualize our mock process, we present images of five TNG100 galaxies in Fig.~\ref{fig:images}.  These galaxies were selected at random within a set of criteria that showcase the significance of the mock process by design.  Each frame covers the full spatial extent of baryons that contribute to the inherent properties of the galaxies (Section \ref{ssec:tng}).  The \Htwo~structure of each galaxy (in its inherent orientation) is shown in the second row of images.  
The code to create the gas images has been placed in a publicly available function.\footnote{See the {\tt build\_gas\_image\_array()} function at \url{https://github.com/arhstevens/Dirty-AstroPy/blob/master/galprops/galplot.py}.}
Overlaid are a dashed ellipse and a solid circle; the former is the inclined two-dimensional \Htwo~half-mass radius, the latter the physical beam size applied in the mock.  We re-image the \Htwo~after applying the mock beam-response weights to each gas cell in the bottom panels. Comparison of the pairs of \Htwo~images highlights the gas that is `missed' by the mock beam.  
We emphasize that these images are purely meant to help visualize the mock process, but images are in no way actually involved in the mock process.  In effect, the beam `sees' a single pixel with an intensity equal to the integral of the images in the bottom panels.
Extra panels are provided in Fig.~\ref{fig:images} for context, including images of stellar content, and plots for their stellar mass, \Htwo~fraction, and mock-to-inherent \Htwo~mass ratio.  Fig.~\ref{fig:images} is discussed further in Section \ref{sec:comparison}.

The mock-observed \Htwo~properties of TNG100 galaxies are only compared to the uncorrected \Htwo~masses from xCOLD GASS.  We discuss the relative merits of this approach versus trusting the beam corrections and comparing to the inherent TNG100 properties at length in Section \ref{sec:comparison}.

% ============================================================ %

\subsubsection{Other properties}
\label{ssec:other}

When presenting `mock' stellar masses for TNG100 galaxies, we approximately account for expected (random) observational uncertainties.  In \citetalias{stevens19}, we calculated this by adding a random number drawn from a Gaussian distribution of standard deviation 0.08\,dex to each inherent logarithmic stellar mass.  In retrospect, this was an underestimate of the typical observational $m_*$ uncertainty.  Even with an advanced SED-fitting code, \citet{robo20} highlight that 0.1\,dex is a ``very optimistic lower limit'' on the true uncertainty of an SED-derived stellar mass.  For this work, we have revised the induced uncertainty on mock stellar masses to 0.2\,dex (consistent with e.g.~\citealt{donnari19}).\footnote{Because our sample has a hard cut of $m_*\!\geq\!10^9\,{\rm M}_\odot$ for \emph{subhaloes}, we \emph{reflect} induced uncertainties at $10^9\,{\rm M}_\odot$ to avoid (i) throwing galaxies away and (ii) artificially over-diluting low-mass bins of actual low-mass galaxies.  For example, if a galaxy with inherent $m_* \! = \! 10^{9.1}\,{\rm M}_\odot$ draws a value of $-0.15$\,dex for its uncertainty, its mock stellar mass is set to $10^{9.05}\,{\rm M}_\odot$ rather than $10^{8.95}\,{\rm M}_\odot$.}

We note that these new mock stellar masses were calculated after the mock redshifts had already been drawn for each galaxy.  This is part of the reason why the redshift distribution at fixed stellar mass for centrals does not align perfectly with xGASS-CO in Fig.~\ref{fig:zdist} (but it is still close).  While this departure from \citetalias{stevens19} means there is greater horizontal smoothing in figures where $m_*$ is on the $x$-axis, we have tested that it is not significant enough to affect any of the discussion or conclusions of this paper.
We have also tested that using a fixed-aperture stellar mass (such as the commonly adopted 30-kpc radius -- see e.g.~section 5.1.1 of \citealt{schaye15} for its relevance) before applying this uncertainty makes little difference to our results.\footnote{The typical inherent aperture radius exceeds 30 kpc for TNG100 galaxies with $m_* \! \gtrsim \! 10^{10.5}\,{\rm M}_\odot$, meaning there would be a decrease in the (mock) stellar masses of these galaxies with the application of a 30-kpc aperture, but this turns out to be irrelevant for this paper.  An extensive comparison of the relevant apertures (and others) is explored in \citet{stevens14}.}

For mock halo masses, we mimic an abundance-matching procedure. That is, we sum the mock stellar masses of galaxies within each host halo to calculate a total stellar mass for that halo, then reassign halo virial masses such that each halo has the same rank when ordered by either mass.  When presenting satellites and centrals for TNG100 in Fig.~\ref{fig:zdist}, we treat the galaxy with the highest mock stellar mass in each halo as the central.  For all other plots, the description in Section \ref{ssec:contam} is followed.

Mock SFRs are designed to average over the historical time-scale that observational SFRs would.  First, we calculate the 20-Myr historically averaged SFR for each galaxy, based on the birth mass and birth time of each star particle.  We calculate a mock sSFR by taking the ratio of this historical SFR to the mock stellar mass, including the induced uncertainty on the latter.
We then add a random Gaussian error of 0.013\,dex on each sSFR, which is typical of the sSFR uncertainty given in the SDSS database for star-forming galaxies in our subsample described in Section \ref{ssec:sdss}.  
For returned sSFRs less than $10^{-11}\,{\rm yr}^{-1}$, the SFR is remeasured on a 1-Gyr time-scale, and a random Gaussian error of 0.025\,dex is then applied to the new sSFR (typical of those given for non-star-forming systems in the SDSS subsample).  Our induced errors here are another departure from the mock procedure in \citetalias{stevens19}, which better serve the comparisons in this paper.
Even after doing this, 17.4\% of TNG100 galaxies still have ${\rm SFR} \! = \! 0$.  Were these observed in a survey like SDSS, their SFRs would unlikely be recorded as zero, though.
In the spirit of finding the `best match', we fit a power law to the passive population of SDSS, measure its scatter (assumed to be Gaussian), and then assign random SFRs drawn from this sequence to the TNG galaxies that still had an SFR lower than the non-zero minimum in the SDSS sample described in Section \ref{ssec:sdss}. More information is provided in Appendix \ref{app:plane}.

No attempt is made to mock the SDSS fibre aperture for SFRs here.  We have instead opted to take the SDSS aperture corrections at face value.  The reason for the difference in philosophy versus our H$_2$ mock measurements lies in the physical scales involved.  The SDSS fibre is 3\,arcsec in diameter.  At $z\!=\!0.05$\,---\,the highest redshift in the interval we compare against\,---\,this translates to $\sim$3\,kpc, i.e.~a radius of $\sim$1.5\,kpc.  This \emph{maximal}  radius is only twice the stellar gravitational softening scale used in TNG100, and is smaller than the smallest beam size used for H$_2$ mock measurements (cf.~Fig.~\ref{fig:zdist}). 
While fibre-like quantities have been measured and adopted in previous TNG100 analyses \citep[e.g.][]{nelson18,donnari19}, our choice to avoid them is twofold: (i) those previous works used a larger radius than the maximal 1.5\,kpc that is relevant here, and (ii) arduous resolution tests would be required to trust results at this relatively small scale.
This means there are potentially strong systematic uncertainties in the SDSS data that we have not explicitly account for/modelled.

Finally, mock \HI~masses are calculated very similarly to \Htwo.  The only difference is the beam size, which for \HI~is 3.5\,arcmin, designed to be consistent with Arecibo and therefore xGASS.  The mock redshifts and method for applying the mock beam are otherwise identical to Section \ref{ssec:H2mock}.

% ============================================================ %

\subsubsection{Cross-contamination of centrals and satellites}
\label{ssec:contam}

As is often the case with comparisons between simulations and observations, even when mock observations for the former are done, typically not all the effects or assumptions that go into the quantities derived from observations are mimicked in the simulation properties.  For example, the central/satellite assignment in the xGASS database \citep[which originates from the group catalogue of][]{yang07} is not pure; that is, many galaxies tagged as satellites will actually be centrals, and vice versa.  
\citet{sb17} have shown that the cross-contamination of satellites and centrals in the \citet{yang07} catalogue can potentially fully explain why the mean (stacked) \HI~fractions of satellites and centrals ($\forall m_* \! > \! 10^9\,{\rm M}_{\odot}$) in the ALFALFA survey \citep{giovanelli05} are so similar, despite a greater difference being predicted by cosmological models of galaxy formation (specifically \ds~in that case).\footnote{\citetalias{stevens19} showed that this problem can be also be solved by mock-observing simulated galaxies (from TNG100) to account for the sky area considered in the ALFALFA all-sky data cube for each galaxy when the stack was made. But, evidently, the equivalent xCOLD GASS mock observations do not offer a solution for \Htwo~here.}
Relatedly, \citet{davies19} have also shown that group finders with less cross-contamination lead to stronger differences between the quenched fractions of centrals and satellites.

When presenting `mock' results for centrals and satellites (with the exception of Fig.~\ref{fig:zdist}), we intentionally cross-contaminate the two populations in TNG100.  
\citet{bravo20} have shown that assuming a fractional contamination between centrals and satellites closely mimics the procedure of running a group catalogue in a simulated galaxy lightcone and assessing contamination there.
In a similar vein, we use weights when calculating statistical quantities (means, medians, and other percentiles) for TNG100 satellites or centrals, which are assigned to ensure we get back a desired effective purity.  We fold in a dependence on host halo mass to these purities, loosely drawn from fig.~6 of \citet{campbell15} for the \citet{yang07} group finder.  These are summarized in Table \ref{tab:weights}.  Additionally, we apply these weights for \emph{each} stellar-mass bin.  For example, when calculating the running statistics for centrals, we first isolate all galaxies with stellar masses in the bin being considered, then break those centrals into halo masses, assign the appropriate weight to each central ($w_{{\rm cen},i}$:~the number from the relevant column in the `centrals' row of Table \ref{tab:weights}), and finally calculate the weights for satellites as
\begin{equation}
\label{eq:weights}
w_{\rm sat} = \frac{1}{N_{\rm sat}} \sum_{i=1}^{4} (1 - w_{{\rm cen},i}) N_{{\rm cen},i}\,.
\end{equation}
Here, $N_{{\rm cen},i}$ refers to the number of centrals in that stellar- \emph{and} halo-mass bin, while $N_{\rm sat}$ refers to the \emph{total} number of satellites in the stellar-mass bin (irrespective of their parent halo mass).  This effectively means that centrals of a given halo mass are contaminated with satellites of all parent halo masses.  This same process is repeated for every stellar-mass bin. All steps are then done again to calculate the running statistics of satellites (swapping `cen' and `sat' in Equation \ref{eq:weights}).%
\footnote{As a point of clarification for the bottom-left panel of Fig.~\ref{fig:sSFR_satcen}, satellites in a given halo mass bin are contaminated with centrals of all halo masses (of the same stellar mass), but there is no contamination from \emph{other} satellites in haloes of different masses.  This is in line with the contamination procedure used in \citet{sb17}.}

\begin{table}
    \centering
    \begin{tabular}{l | c c c c}\hline
        $\log_{10}\!\left(M_{\rm 200c}^{\rm parent}/{\rm M}_\odot\right)$ & $<\!12$ & $[12,13)$ & $[13,14)$ & $>\!14$ \\ \hline
        Central purity & 0.95 & 0.90 & 0.80 & 0.70 \\
        Satellite purity& 0.50 & 0.65 & 0.70 & 0.75 \\ \hline
    \end{tabular}
    \caption{Purities assumed when artificially cross-contaminating centrals and satellites for mock TNG results.  These vary with galaxy type (central or satellite:~rows) and parent halo mass ($M_{\rm 200c}^{\rm parent}$:~columns). Cf.~\citet{campbell15}.}
    \label{tab:weights}
\end{table}

This contamination process is another feature that was not included in the mock results of \citetalias{stevens19}.
Because there are several new features to the mock process in this paper, we update the key xGASS-related results from \citetalias{stevens19} in Appendix \ref{app:HI} to be consistent with the methods described here.

% ============================================================ %

\subsection{Conventions}

Throughout this paper, we adhere to the following conventions.
\begin{itemize}
\item Whenever we refer to an `\Htwo~fraction', we exclusively mean the ratio of \Htwo~mass to stellar mass of a galaxy (and likewise for `\HI~fractions').
\item All cosmology-dependent quantities\,---\,both simulated and observational\,---\,assume (or have been adjusted to assume) $\Omega_m \! = \! 0.3089$, $\Omega_{\Lambda} \! = \! 0.6911$, and $h \! = \! 0.6774$ \citep{planck16}.
\item Likewise, all IMF-dependent quantities assume a \citet{chabrier03} IMF.  When converting from a \citet{kroupa01} IMF, we assume $m_{\rm *,Chabrier} \! = \! \frac{0.61}{0.66} m_{\rm *,Kroupa}$ and ${\rm SFR}_{\rm Chabrier} \! = \! \frac{0.63}{0.67}\,{\rm SFR}_{\rm Kroupa}$ \citep[see][]{madau14}.
\item Zeroes are never discarded when calculating statistical quantities such as means, medians, and other percentiles.
\item Logarithms are always taken \emph{after} the statistical quantity of interest is computed.
\item Upper-case $R$ refers to three-dimensional/spherical radii.  Lower-case $r$ refers to two-dimensional/cylindrical radii.  
\item Upper-case $M$ (or $\mathcal{M}$) refers to the mass of haloes.  Lower-case $m$ refers to the mass of galaxies.
\item When binning data, we use bins of nominal (but not strict) width 0.2\,dex. The absolute least number of galaxies we use in a bin is 15.  No galaxies are discarded when binning.  When necessary, bin widths automatically adapt to ensure both latter criteria are fulfilled.
\item There are no `helium contributions' to \HI~or \Htwo~masses.
\item Whenever we refer to `halo mass', we always mean that of the `host' or `parent' halo.  We use these terms interchangeably.
\end{itemize}

% ============================================================ %
% ============================================================ %
% ============================================================ %
% ============================================================ %
% ============================================================ %
% ============================================================ %
% ============================================================ %

\section{A census of molecular gas in galaxies at redshift zero}
\label{sec:comparison}

Before addressing effects of environment, let us devote this section to setting the scene for what the xCOLD GASS survey and TNG100 simulation tell us about the \Htwo~properties of galaxies at \zo~on the whole (and in a comparative sense).

\subsection{H$_2$ relative to stellar content}
\label{ssec:H2frac}

\begin{figure}
\centering
\includegraphics[width=0.95\textwidth]{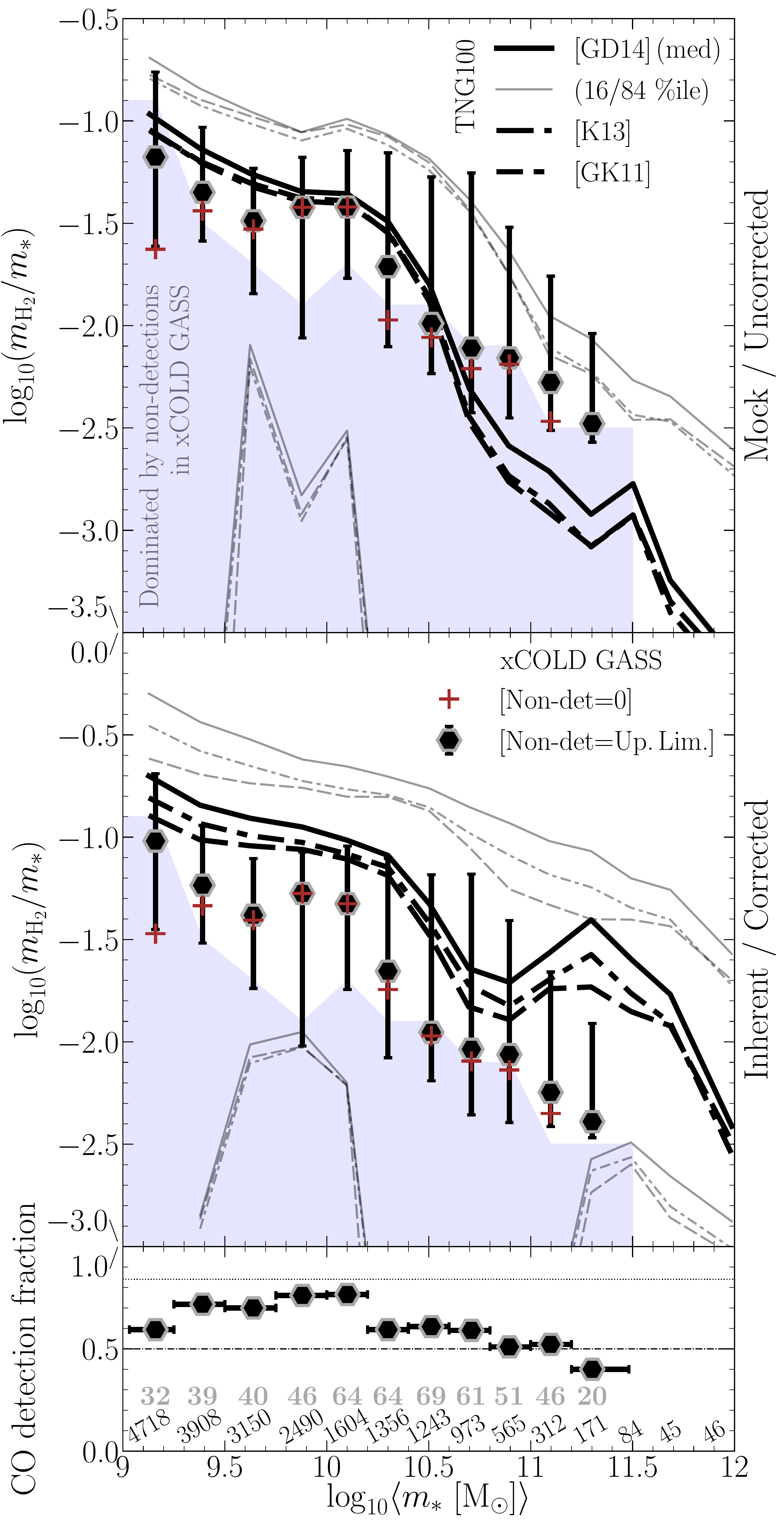}
\caption{Molecular-hydrogen fraction of all galaxies (i.e.~centrals and satellites together) as a function of stellar mass. {\bf The top panel} compares mock-observed properties from TNG100 with the data from xCOLD GASS that have \emph{not} been beam-corrected.  {\bf The second panel} compares the inherent properties of TNG100 galaxies with the beam-corrected xCOLD GASS values. In each case, the running 16th, 50th, and 84th percentiles of both the simulation and survey data are given by lines and hexagons+error bars, respectively.  Hexagons represent medians when non-detections are assumed to take their upper-limit value, whereas the plus symbols represent medians when non-detections are set to have $m_{\rm H_2}\!=\!0$.  
To mitigate redundancy, we exclude a second set of error bars for the plusses; these would extend from the top of the black error bars down to $-\infty$.
The different line styles for TNG100 correspond to different \HI/\Htwo~prescriptions.  
%For both the simulation and observed data, bins of minimum width 0.2\,dex (0.25\,dex for $m_*\!<\!10^{10}\,{\rm M}_{\odot}$) with a minimum of 20 galaxies were used.  
%The precise $x$-axis position for each bin as plotted is the mean stellar mass for galaxies in that bin.  
The shaded region in these panels covers where the observations are dominated by non-detections.  
{\bf The bottom panel} shows the detection fraction of xCOLD GASS galaxies, with horizontal error bars covering the full bin width. 
Printed numbers are the galaxy counts in each bin for xCOLD GASS (grey, bold) and TNG100 (black, angled).  
For reference, dotted and dot-dashed horizontal lines are respectively given at detection fractions of 0.84 and 0.5.}
\label{fig:H2Frac}
\end{figure}

In the top two panels of Fig.~\ref{fig:H2Frac}, we show the variation in the \Htwo~fraction of galaxies from both TNG100 and xCOLD GASS as a function of stellar mass, including the relation's scatter (cf.~fig.~3 of \citealt{diemer19}).  The top panel makes no correction to the CO observations for the finite beam, and instead applies the same beam to the simulations to derive comparable `mock observed' \Htwo~masses (as outlined in Section \ref{ssec:mock}).  Forward-modelled uncertainties on stellar masses are also included.  By contrast, the second panel \emph{does} employ the \citet{saintonge12} beam correction to the observations, while the inherent (gravitationally bound) properties of TNG galaxies are plotted.  In both cases, non-detections in xCOLD GASS assume their upper-limit measurements for the percentiles shown with hexagons and error bars, and are instead set to zero for the medians shown by the plusses.
Loosely, both datasets in both panels suggest that higher-mass galaxies have lower \Htwo~fractions, with significant scatter at all masses.
It is important that we show both these panels, as it highlights the difficulty and care required to compare simulations with observations; the level of agreement is notably different between the two.  At the low-mass end, the uncorrected and mock quantities seem better matched. But at the high-mass end, the choice of comparison either leads to the conclusion that TNG100 galaxies are systematically too gas-rich or contrarily that they are \emph{suggestively} too gas-poor\footnote{The detection fraction of xCOLD GASS is too low to claim this definitively.} (modulo \emph{systematic} uncertainties in xCOLD GASS masses, which are not shown).

Fig.~\ref{fig:images} provides a deeper understanding of why there can be such significant differences between the mock and inherent \Htwo~properties of some TNG100 galaxies, even while other galaxies show negligible differences.
The first two columns of images show two galaxies with comparable stellar masses and \Htwo~fractions, yet vastly different \Htwo~structure. If the first galaxy were observed face-on, its two-dimensional \Htwo~half-mass radius would be measured as $r_{\rm H_2}^{\rm half} \! = \! 2.2$\,kpc, well within the mock beam size.  In fact, nearly all the \Htwo~in this galaxy fits within the mock beam.  By contrast, the second galaxy has $r_{\rm H_2}^{\rm half} \! = \! 7.3$\,kpc, which exceeds the beam size.  The outer half of its \Htwo~mass also lies in extended spiral arms, all of which is missed by the beam.
The next three galaxies all have higher stellar masses, each closer to $10^{11}\,{\rm M}_{\odot}$.  Again, these highlight a variety of situations where anywhere from 35\% to 88\% of the galaxies' \Htwo~is missed by the mock beam.  The most extreme of these three examples is actually representative of the \emph{average} galaxy at this stellar mass, as discerned from the top-middle panel of Fig.~\ref{fig:images}.

All of this motivates the question: is one form of comparison (mock/uncorrected or inherent/corrected) more meaningful than the other?
An argument can be made that, because the beam correction for xCOLD GASS requires a model for how \Htwo~in galaxies should be distributed (i.e. that it is distributed exponentially, with a half-mass radius equal to the radius enclosing half the galaxy's star formation rate), the mock/uncorrected comparison is the most direct.  Unfortunately, this implicitly relies on \Htwo~being distributed in TNG galaxies the same way it is in reality; otherwise the applied beam would exclude a different fraction of the true \Htwo~mass of the galaxy.  As it happens, \citet[][see their fig.~7]{diemer19} have shown that \Htwo~in TNG galaxies can extend to higher radii than found in observations \citep[cf.][]{bolatto17}.  While the difference in extent is only of order tens of percent on average, there is a tail of galaxies that stretches out to factors of a few.
We have since learned that higher-mass galaxies lie in this tail at greater frequency.
Fig.~\ref{fig:images} tells a consistent story, where several examples of galaxies with \Htwo~half-mass radii notably exceeding their stellar half-mass radii are shown.  
Based on this then, if the goal is to compare the integrated \Htwo~masses of galaxies, one can make a counter-argument that the inherent/corrected comparison might be more meaningful.

\begin{figure}
\centering
\includegraphics[width=0.956\textwidth]{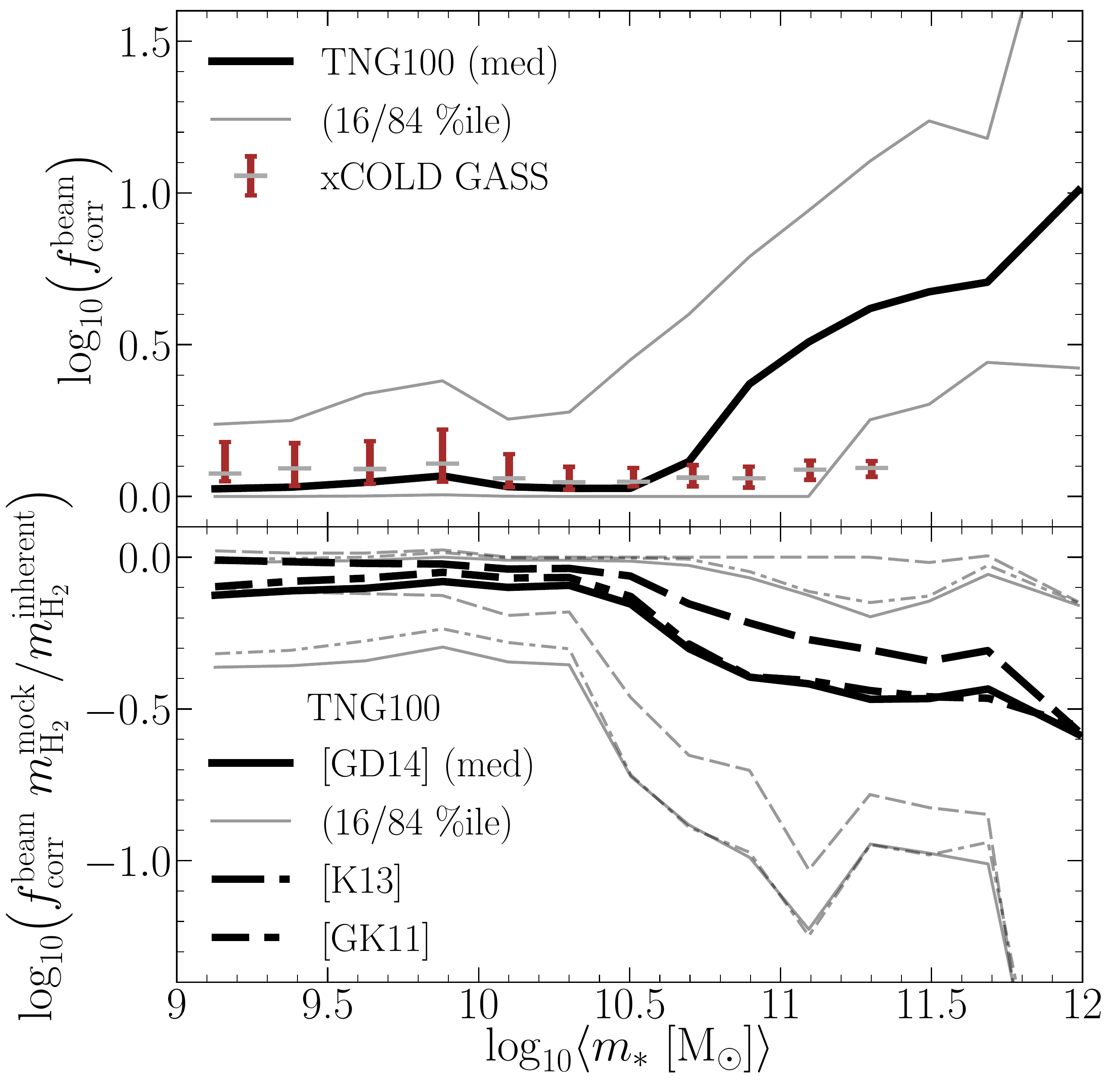}
\caption{{\bf Top panel:} Running distributions of beam correction factors for xCOLD GASS and equivalents for TNG100, per Equation (\ref{eq:corr}), as a function of stellar mass (inherent for TNG100).  Horizontal dashes for xCOLD GASS are medians, while error bars cover the 16th--84th percentile range.  $f^{\rm beam}_{\rm corr}$ has no dependence on $m_{\rm H_2}$, and thus all detections and non-detections for xCOLD GASS are treated equivalently here, plus there is no \HI/\Htwo~prescription dependence for TNG.
{\bf Bottom panel:} Ratio of TNG100 effective beam correction factors to the actual difference factor between the inherent and mock \Htwo~masses.  Running medians, 16th and 84th percentiles are shown for each \HI/\Htwo~prescription.  If Equation (\ref{eq:corr}) were consistently accurate, all lines in this panel should be close to 0.  Instead, $f^{\rm beam}_{\rm corr}$ is seen to underestimate the `true' correction factor for a significant fraction of galaxies.  The $f^{\rm beam}_{\rm corr}$ quantities for TNG100 galaxies are not used anywhere else in this paper.  Galaxies with ${\rm SFR} \! = \! 0$ \emph{are} included in this plot; they are interpreted to have $f^{\rm beam}_{\rm corr} \! = \! m_{\rm H_2}^{\rm inherent} / m_{\rm H_2}^{\rm mock} \! = 1$.}
\label{fig:corr}
\end{figure}

The truth is that neither form of comparison is perfect.  But there is a test we can perform to help inform when one comparison might be more meaningful than the other.  Namely, we can calculate an `effective beam correction factor' for each TNG100 galaxy using Equation (\ref{eq:corr}), and compare the distribution of these factors with (i) those of xCOLD GASS, and (ii) the ratio of inherent to mock \Htwo~masses of TNG100.  Importantly here, Equation (\ref{eq:corr}) is \emph{independent} of any actual \Htwo~properties of the galaxies.  We define $r_{\rm SFR}^{\rm half}$ for TNG galaxies using the instantaneous SFRs of their gas cells.  If Equation (\ref{eq:corr}) is an accurate method for beam correction, then $f^{\rm beam}_{\rm corr}$ should be almost the same as $m_{\rm H_2}^{\rm inherent} / m_{\rm H_2}^{\rm mock}$ for each TNG galaxy.  To check this, we plot the running distributions of both $f^{\rm beam}_{\rm corr}$ and the ratio of it to $m_{\rm H_2}^{\rm inherent} / m_{\rm H_2}^{\rm mock}$ in Fig.~\ref{fig:corr} (both quantities are interpreted to be equal to 1 for galaxies with no star formation activity).

For $m_* \! \lesssim 10^{10.3}\,{\rm M}_{\odot}$, the bottom panel of Fig.~\ref{fig:corr} highlights that Equation (\ref{eq:corr}) does a good job of predicting $m_{\rm H_2}^{\rm inherent} / m_{\rm H_2}^{\rm mock}$ for the median TNG galaxy.  However, the 16th percentiles show that $f^{\rm beam}_{\rm corr}$ can underestimate this ratio by a factor of $\sim$2 at significant frequency.  
This notably depends on the \HI/\Htwo~prescription; in the case of \citetalias{k13}, \Htwo~does not extend as far out for each galaxy, and thus $f^{\rm beam}_{\rm corr}$ remains accurate within $\sim$30\% (still as an underestimate) for most galaxies.\footnote{As a point of comparison, the uncertainty taken for beam corrections in \citet[][section 2.5]{saintonge17} is 15\%.}  
For the same mass range, the top panel of Fig.~\ref{fig:corr} shows that the running distributions of $f^{\rm beam}_{\rm corr}$ for TNG100 and xCOLD GASS are fairly similar.  In principle, it does not matter if these are quantitatively identical or not; differences between distributions here are the result of differences in the distributions of $r_{\rm SFR}^{\rm half}$:\footnote{Per Equation (\ref{eq:corr}), the only other thing that could matter is differences in the distribution of $\sigma_{\rm beam}$ in physical units, but these are the same by construction.} the point of the correction being that it accounts for this.  This comparison simply tells us that the SFR sizes\,---\,and therefore the gas sizes\,---\,of TNG100 galaxies are physically reasonable for the majority with $m_* \! \lesssim 10^{10.3}\,{\rm M}_{\odot}$.  Given this, and the unsurprising fact that $f^{\rm beam}_{\rm corr}$ is not identical to $m_{\rm H_2}^{\rm inherent} / m_{\rm H_2}^{\rm mock}$, forward-modelling the beam is, objectively, a better way to compare simulations with observations.  In other words, the mock/uncorrected comparison should be more meaningful than the inherent/corrected comparison on the basis that it convolves a known gas distribution with a known beam to compare to a directly observed quantity, as opposed to \emph{assuming} how gas is distributed in xCOLD GASS galaxies to effectively deconvolve the beam.
The fact that we find $f^{\rm beam}_{\rm corr}$ to frequently underestimate the beam correction factor explains why TNG100 and xCOLD GASS are better aligned at low masses in the top panel of Fig.~\ref{fig:H2Frac} than in the middle panel.  

However, the situation is not so clear-cut at higher masses.  The top panel of Fig.~\ref{fig:corr} shows wild differences in $f^{\rm beam}_{\rm corr}$ between TNG100 and xCOLD GASS for $m_* \! \gtrsim 10^{10.7}\,{\rm M}_{\odot}$.  This is most likely symptomatic of gas being unrealistically distributed in the simulated galaxies.
Moreover, as seen in the bottom panel of Fig.~\ref{fig:H2Frac}, $f^{\rm beam}_{\rm corr}$ and $m_{\rm H_2}^{\rm inherent} / m_{\rm H_2}^{\rm mock}$ start diverging at $m_*\!>\!10^{10.3}\,{\rm M}_{\odot}$.  This implies that the radial gas profiles of these galaxies deviate significantly\,---\,i.e.~more than one would expect\,---\,from a symmetric exponential.  The major differences in the mock and inherent \Htwo~masses seen at these stellar masses are therefore somewhat artificial (or, in other words, not predictive of reality).  In light of this, we suggest that the inherent/corrected comparison with xCOLD GASS is a safer choice than the mock/uncorrected comparison at high stellar masses, albeit still an imperfect one.

Given the uncertainties raised above, our philosophy is to present both the mock/uncorrected and inherent/corrected comparisons where relevant throughout this paper.  Each requires a grain of salt in its interpretation, but both are needed to provide an as-complete-as-possible picture.

As shown explicitly in the bottom panel of Fig.~\ref{fig:H2Frac}, a significant number of the xCOLD GASS galaxies are not detected in CO; in fact, for the highest-mass bin, the majority are non-detections.  This is also reflected in the shaded regions in the top two panels; these cover the area where there are more upper limits in the observational data than detections.%
\footnote{While the survey was designed with an \Htwo~fraction detection limit (mentioned in Section \ref{ssec:xcold}), in practice, some non-detections have higher upper limits on their \Htwo~fractions than other detections' measurements at the same stellar mass.}
Because the detection fraction of xCOLD GASS galaxies is $<\!0.84$ in all mass bins, we cannot show the 16th percentile in Fig.~\ref{fig:H2Frac} for xCOLD GASS \Htwo~fractions if non-detections are assumed to have no \Htwo~(those percentiles would all be zero, i.e.~$-\infty$ in log space).  In essence, this means the lower half of the scatter in the \Htwo~fraction--stellar mass relation is observationally unconstrained.  We therefore cannot make meaningful statements about how representative the frequency of properly \Htwo-poor galaxies in TNG is of reality.

Nevertheless, we note that the 16th percentile of inherent \Htwo~fractions of TNG100 galaxies shows a dip over $10^{10.2}\!<\!m_*/{\rm M}_{\odot}\!<\!10^{11.3}$, reaching a trough outside the axes of Fig.~\ref{fig:H2Frac}.
This lines up with the dip in \HI~fractions identified in \citetalias{stevens19}.  This feature effectively disappears in the mock properties, in part because the \Htwo~fractions of all the most massive galaxies drop after the application of the relatively small beam.  As discussed in \citetalias{stevens19}, this feature is most likely related to the onset of kinetic feedback from massive black holes / active galactic nuclei \citep[AGN\,---\,for details on the implementation and result of said feedback, see][respectively]{weinberger17,zinger20}.

In general, there is a slightly greater sensitivity to the choice of \HI/\Htwo~prescription for TNG \Htwo~results than for \HI~(cf.~\citealt{diemer18,diemer19}; \citetalias{stevens19}).  This is perhaps unsurprising, as \HI~tends to dominate over \Htwo, meaning small fractional changes in \HI~between prescriptions are reflected by larger fractional changes in \Htwo~(as the total \HI+\Htwo~is fixed).  The source of this difference is most readily understood by looking at the \mHtwo/\mHI~ratios of the simulated galaxies, which we assess next.

% ============================================================ %

\subsection{H$_2$ relative to \HI~content}

\begin{figure}
\centering
\includegraphics[width=0.937\textwidth]{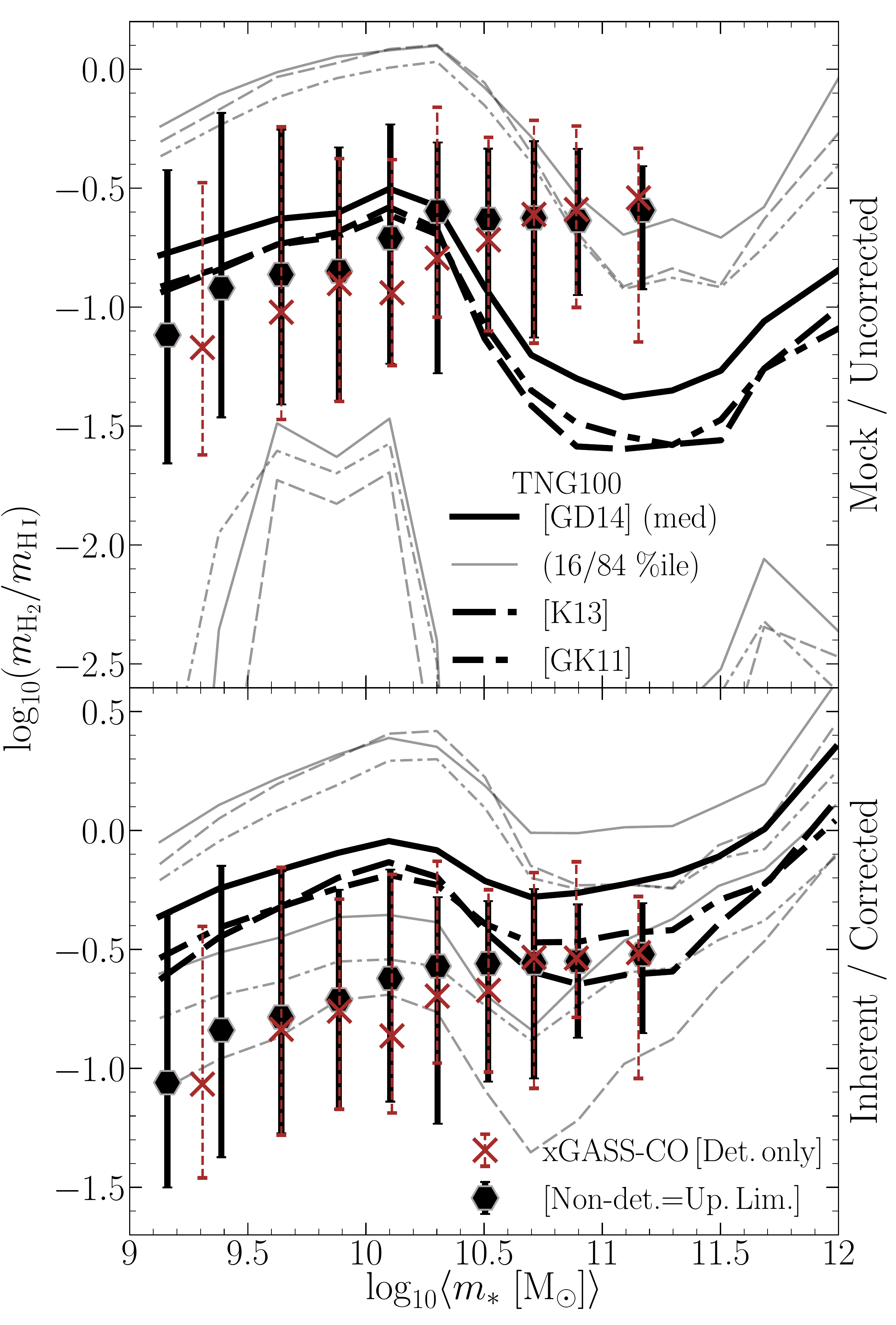}
\caption{Ratio of molecular to atomic hydrogen in galaxies as a function of their stellar mass, at $z\!\simeq\!0$.  Plotting conventions follow Fig.~\ref{fig:H2Frac} for TNG100 data.  For xGASS-CO data, filled symbols (with solid error bars) give the median (plus 16th \& 84th percentiles) for the complete overlap in xGASS and xCOLD GASS, where non-detections in either \HI~or CO assume their upper limits.  Exes with dashed error bars \emph{exclusively} use the galaxies detected in \emph{both} \HI~and CO.}
\label{fig:H2HIFrac}
\end{figure}

In Fig.~\ref{fig:H2HIFrac}, we show the \Htwo/\HI~mass ratios of all TNG100 and xGASS-CO galaxies as a function of stellar mass.  As documented by \citet{catinella18}, the \Htwo/\HI~mass ratios of xGASS-CO galaxies rise with stellar mass on average.  But even over 2.5\,dex in stellar mass, this rise is less than (or at least comparable to) the relation's scatter.  We again show two panels that compare the mock-observed simulation properties with uncorrected observations (top) and inherent simulation properties with corrected observations (bottom).  Although, we note that no beam correction has been applied to the \HI~content of xGASS-CO galaxies in the bottom panel of Fig.~\ref{fig:H2HIFrac} (it has only been applied for \Htwo, as given in the public xGASS and xCOLD GASS databases); 
see \citetalias{stevens19} for the potential significance of this (much larger) beam.  
We also need to be particularly careful about the treatment of non-detections; while some galaxies are detected in both \HI~and CO, some are only detected in one, some the other, and some neither.  We still present running percentiles where non-detections (in both phases) are set to their upper limits.  But it no longer makes sense to plot the equivalent when non-detections are set to zero; to do this for \mHI~would mean dividing by zero (for the same reason, only TNG100 galaxies with $m_{\rm H\,{\LARGE{\textsc i}}} \! > \! 0$ are included in Fig.~\ref{fig:H2HIFrac}).  Instead, we show a second set of points (exes) with dashed error bars, which exclusively consider galaxies detected in \emph{both} phases.  This again provides two imperfect means of comparing to simulation results.  Somewhat reassuringly though, there is little difference between the two sets of observational points.  

TNG100 mock results are broadly consistent with observations in the left half of the top panel of Fig.~\ref{fig:H2HIFrac}, but are again hampered by the overextension of \Htwo~at high stellar masses in the simulation. Here, the \Htwo/\HI~mass ratios show a systematic drop from $m_*\!\gtrsim\!10^{10.3}\,{\rm M}_\odot$, before rising again at $m_*\!\gtrsim\!10^{11}\,{\rm M}_\odot$.  While the depth of this feature is tied to our mock procedure, comparison with the bottom panel of Fig.~\ref{fig:H2HIFrac} shows that the feature still exists for the inherent galaxy properties, just more weakly.  As such, there must be an underlying physical cause for this feature in the simulation.
This suggests that gas from the galaxies' centres\,---\,i.e.~where the majority of \Htwo~but only a fraction of \HI~is \citep[see e.g.][]{leroy08,ob09,stevens19b}\,---\,is being removed or depleted.
Based on differences in mock/inherent \HI~masses alone, especially when contrasting satellites and centrals, \citetalias{stevens19} hypothesized that AGN feedback was likely responsible for displacing gas from galaxies' centres in TNG100 at the mass scale in question.  Having imaged several galaxies, we can confirm there are clear holes and asymmetries in the gas structure of many galaxies at these masses, as would be expected from ejective feedback at each galaxy's centre \citep[for a related analysis of galactic outflows in the TNG model, see][]{nelson19b}.  The fourth galaxy in Fig.~\ref{fig:images} is a prime example of this.  Most of the \Htwo~on one side of the galaxy has been blown out, with the other side clearly disrupted, as seen by its lack of alignment with the stellar disc.  This physical effect is then compounded when mock measurements are made, as it is then \emph{only} the central region of the galaxy that would contribute meaningfully to the \Htwo~mass, but that gas has been displaced. 

While we cannot claim that TNG100 is a 100\% faithful representation of reality when it comes to galaxies' \Htwo~content\,---\,indeed, we have identified some issues\,---\,Figs~\ref{fig:H2Frac} \& \ref{fig:H2HIFrac} at least give us confidence that the simulation is `in the right ball park', so to speak (especially for $10^9 \! \leq \! m_*/{\rm M}_{\odot} \! \lesssim \! 10^{10.2}$).  Similar statements could be made about the EAGLE\footnote{Evolution and Assembly of GaLaxies and their Environments} simulations (cf.~fig.~5 of \citealt{lagos15b}; also see \citealt{dave20}).  It will likely be years before next-generation simulations and CO surveys will allow us to make comparative statements that are more grand and robust than this.  In the interim, the simulations that we have on hand still provide worthwhile synthetic laboratories to study the effects of environment on \Htwo, bearing in mind the caveats that carry through from the above.

% ============================================================ %
% ============================================================ %
% ============================================================ %
% ============================================================ %
% ============================================================ %
% ============================================================ %

\section{Molecular gas in galaxies by environment}
\label{sec:env}

\subsection{Central galaxies versus satellites}
\label{ssec:satcen}

\begin{figure*}
\centering
\includegraphics[width=0.73\textwidth]{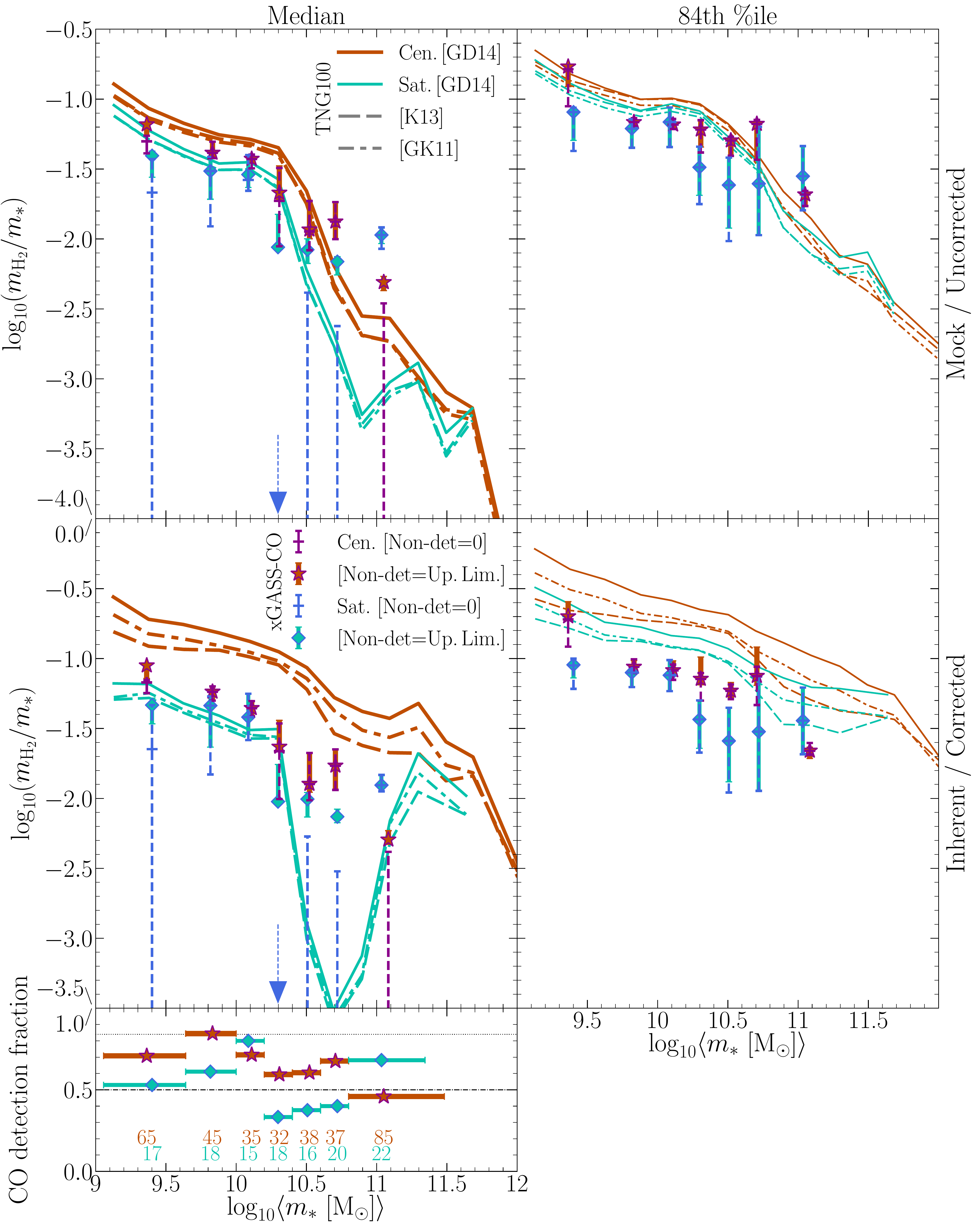}
\caption{Molecular-hydrogen fraction as a function of stellar mass for central and satellite galaxies presented separately.  The layout of this figure differs from Fig.~\ref{fig:H2Frac}.  Here, the left-hand panels only show the median trend for TNG100 and xGASS-CO, while the right-hand panels show the running 84th percentile.  
Error bars for xGASS-CO percentiles are 68\% confidence intervals \emph{of those percentiles} from bootstrapping\,---\,points accompanying the error bars still represent the raw percentiles from the data.  
Two sets of error bars exist in each panel: one for when we assume that non-detections take their upper limit value (solid bars with either pentagram or diamond points), and one for when we assume non-detections have zero \Htwo~(dashed bars with plusses).  Downward arrows are used when the entire 68\% confidence interval falls below the axes.  The same bins are used for both centrals and satellites.  The bottom-left panel shows the detection fraction, width, and number of galaxies in each xGASS-CO bin.}
\label{fig:H2Frac_SatCen}
\end{figure*}

As outlined in Section \ref{sec:intro}, one of the main aims of this paper is to investigate how centrals and satellites differ in their \Htwo~content at fixed stellar mass.  In Fig.~\ref{fig:H2Frac_SatCen}, we show precisely this for xGASS-CO (i.e.~the xCOLD GASS galaxies that have central/satellite status determined as part of xGASS, per Section \ref{ssec:xcold}) and TNG100 at \zo.  Similar to Fig.~\ref{fig:H2Frac}, we compare the mock/uncorrected values in the top panels, and the inherent/corrected values in the middle panels.  Because we are now subsampling the observational data, we are dealing with weaker statistics.  This, combined with there being visually more information in the figure, means we have shown the running medians and 84th percentiles in separate columns.  We exclude a column of panels for the 16th percentile here primarily (but not exclusively\,---\,see below) due to the low detection fraction in the observations.  Importantly, we can trust that the distribution of parent halo masses of satellites in xGASS-CO is very similar to that of TNG100 (Fig.~\ref{fig:zdist}), meaning their gas properties at fixed stellar mass should be directly comparable.

Looking at the observational data alone in any of the main four panels of Fig.~\ref{fig:H2Frac_SatCen}, there is little evidence to support the idea that satellites are \emph{significantly} depleted of their \Htwo~at a particular mass scale.
But some degree of systematic difference does appear to be present, as 5--6 of the seven stellar-mass bins show satellites to have lower median and 84th-percentile \Htwo~fractions than centrals.
While the bootstrapping errors on these percentiles imply this is not statistically significant for any singular bin, the likelihood of randomly observing this result for \emph{all} bins simultaneously is low (we elaborate below).
Nevertheless, the results of Fig.~\ref{fig:H2Frac_SatCen} contrast with the directly comparable plot for \HI~in xGASS \citepalias[fig.~5 of][]{stevens19}, where there is a clearer distinction between the distribution of \HI~fractions of satellites and centrals for all fixed $m_*\!\lesssim\!10^{10.75}\,{\rm M}_{\odot}$ (see Appendix \ref{app:HI}).  We have tested that those \HI~results remain when we select xGASS-CO galaxies exclusively, implying that the \emph{relative} lack of observed difference in \Htwo~for centrals and satellites cannot be fully explained away by the subsample being small and/or potentially unrepresentative.

By contrast, satellites in TNG100 show a clear depletion in their \Htwo~fractions relative to centrals of the same stellar mass.  
The inherent results of the simulation find the median satellite to be $\sim$0.6\,dex (a factor of $\sim$4) poorer in \Htwo~than the median central at fixed $m_* \! < \! 10^{10.3}\,{\rm M}_\odot$.
From a theoretical, qualitative stand-point, this is not an unreasonable result.  One can imagine that satellites' molecular gas would continue to be consumed in star formation (and accompanied feedback), but would be unable to efficiently replenish, due to the denial of cosmological gas accretion \citep[similar to the picture outlined by][]{sb17}.  
If the depletion of \Htwo~were to be approximately linear with time, we should see a greater difference between centrals and satellites in log space for low percentiles, and likewise a smaller difference at high percentiles.  Indeed, there is only a 0.2--0.3\,dex difference (a factor of $<$2) in the inherent 84th percentile of centrals and satellites in TNG100 (lower-right panel of Fig.~\ref{fig:H2Frac_SatCen}).
At the same time, for those satellites that experience a pericentric passage close to the halo centre, ram pressure from the relative motion of the intrahalo medium would be so strong that it could entirely remove the interstellar medium (i.e.~including \Htwo).  Another reason we opted to exclude panels for the 16th percentile in Fig.~\ref{fig:H2Frac_SatCen} is that it equals zero ($-\infty$ in log) for TNG100 satellites $\forall m_*\!\geq\!10^9\,{\rm M}_{\odot}$.

It is starkly obvious that the separation between centrals and satellites diminishes significantly when we look at the mock-observed simulation outputs.  The separation between running medians has been reduced from $\sim$0.6 to $\sim$0.2\,dex.  From testing (not shown here), we learned that while the application of the mock beam (Section \ref{ssec:H2mock}) has an impact on decreasing this separation, the most important factor is the cross-contamination of centrals and satellites (Section \ref{ssec:contam}).
That is, if we apply the contamination to the inherent properties, we find a separation $\sim$0.3\,dex.  Similarly, had we excluded the cross-contamination from our mock procedure (but kept the other aspects the same), we would have found a separation of $\sim$0.4\,dex.
Satellite--central cross-contamination is always present in the xGASS-CO data, regardless of whether beam corrections are considered or not.  There is no way to correct for this contamination without applying a completely different group-finding algorithm to SDSS; even then, a `perfect' group finder does not exist to our knowledge, nor might it ever. The best option (arguably, the only option) is to forward-model the contamination, like we do with the mock results of TNG100.  As such, the mock/uncorrected comparison is \emph{always} more meaningful than the inherent/corrected comparison when investigating the \emph{relative} influence of environment, despite the concerns about the \emph{absolute} mock \Htwo~properties at high stellar masses raised in Section \ref{ssec:H2frac}.

\begin{figure}
\centering
\includegraphics[width=0.9838\textwidth]{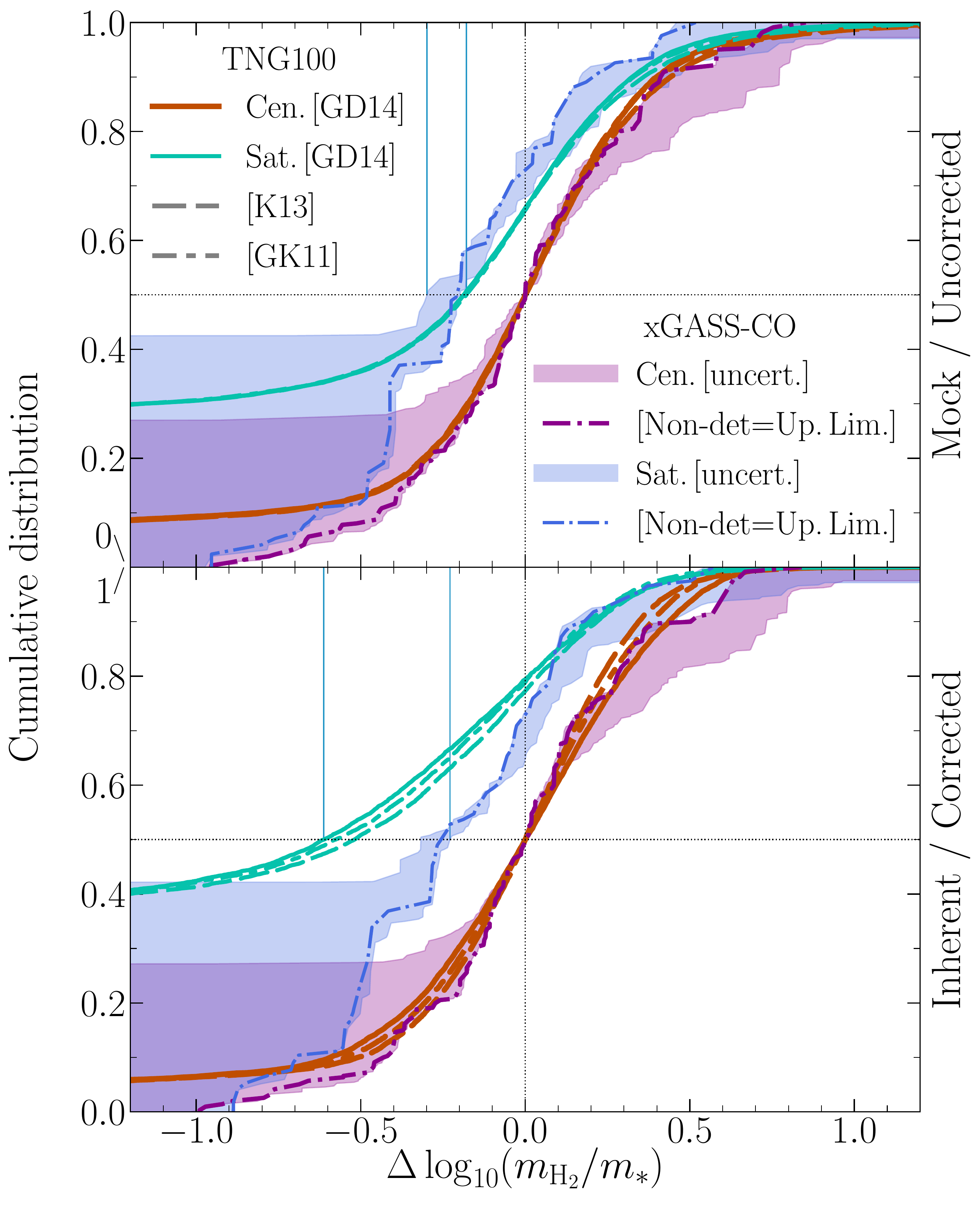}
\caption{Cumulative distributions of the \Htwo~fractions of galaxies \emph{relative} to the median \Htwo~fraction of central galaxies at the same stellar mass (from the same sample; see Equation \ref{eq:Delta}).  Distributions for each \HI/\Htwo~prescription are plotted for TNG100; they are nearly all identical in the top panel.  Dot-dashed curves for xGASS-CO assume the upper limits as the \Htwo~fractions of non-detections.  The shaded regions for xGASS-CO show the uncertainty range, accounting for (i) the possibility that non-detections might have no \Htwo~at all and (ii) the uncertainty on the median \Htwo~fraction of centrals in each stellar-mass bin (as seen in Fig.~\ref{fig:H2Frac_SatCen}).  Each galaxy in xGASS-CO has a weighted contribution based on the relative frequency of galaxies at its stellar mass in the survey to that in TNG100 (Section \ref{ssec:xcold}), where weights for centrals and satellites are calculated independently.  Lower-mass galaxies therefore dominate these distributions (by virtue of the shape of the stellar mass function). Systematic differences between centrals and satellites are readily identified by the horizontal separation of their respective curves where they intercept the horizontal, dotted line; the vertical, solid lines show the range of these intercepts for satellites from both TNG100 and xGASS-CO.  By definition, all curves for centrals intersect at $(0.0,0.5)$, i.e.~where the dotted lines also intersect.  Per the top panel, TNG100 predicts that xGASS-CO should find a systematic difference of $\sim$0.2\,dex between satellites and centrals.  Indeed, xGASS-CO is consistent with this prediction, finding a difference of 0.2--0.3\,dex within uncertainty.  The bottom panel is included for completeness, but\,---\,unlike the top panel\,---\,it does not provide a fair comparison between TNG100 and xGASS-CO, as no accounting for the cross-contamination of centrals and satellites has been made (see Section \ref{ssec:satcen} for details).}
\label{fig:Delta}
\end{figure}

A natural question to ask is:~do the xGASS-CO data favourably support our prediction that the survey should see a 0.2-dex systematic difference between the \Htwo~fractions of centrals and satellites?
Close inspection of the top panels of Fig.~\ref{fig:H2Frac_SatCen} suggest that the data are consistent with this prediction.  But the uncertainties in each stellar-mass bin for xGASS-CO make it difficult to read how quantitatively precise this apparent consistency is.  By combining the data from all bins, however, we can make a more rigorous comparison.  To this end, we calculate the difference in \Htwo~fraction for each galaxy from the median of \emph{centrals} in each bin, and thus define
\begin{multline}
\label{eq:Delta}
\Delta \log_{10}\!\left( m_{\rm H_2}/m_*\right) \equiv \log_{10}\!\left( m_{\rm H_2}/m_*\right) - \\ \log_{10}\!\left( {\rm Median}\!\left[ \frac{m_{\rm H_2}}{m_*} (m_*,{\rm cen.})\right] \right)\,.
\end{multline}
In Fig.~\ref{fig:Delta}, we show the cumulative distributions of this property for both centrals and satellites from each of TNG100 and xGASS-CO. The top panel shows the predicted distributions from TNG100 to lie almost on top of the uncertainty range of xGASS-CO.  
A Kolmogorov--Smirnov (KS) test of the observational data is enough to rule out the possibility that satellites and centrals follow the same distribution in $\Delta \log_{10}\!\left( m_{\rm H_2}/m_*\right)$.\footnote{The two-sided, two-sample KS statistic for the two xGASS-CO dot-dashed curves in the top panel of Fig.~\ref{fig:Delta} is 0.31, with a $p$-value of order $10^{-3}$.}
Not only does this show that the survey is consistent with the prediction of TNG100 regarding the systematic difference between centrals and satellites, but this also means TNG100 predicts \emph{very} similar scatter in \Htwo~fractions to what xGASS-CO observes for both satellites and centrals.
The bottom panel of Fig.~\ref{fig:Delta} highlights how different the `true' distributions of \Htwo~fractions are for TNG100 satellites.  We include this panel for completeness, but remind the reader that the top panel is the definitive comparison to be made with observations.

\begin{figure}
\centering
\includegraphics[width=0.937\textwidth]{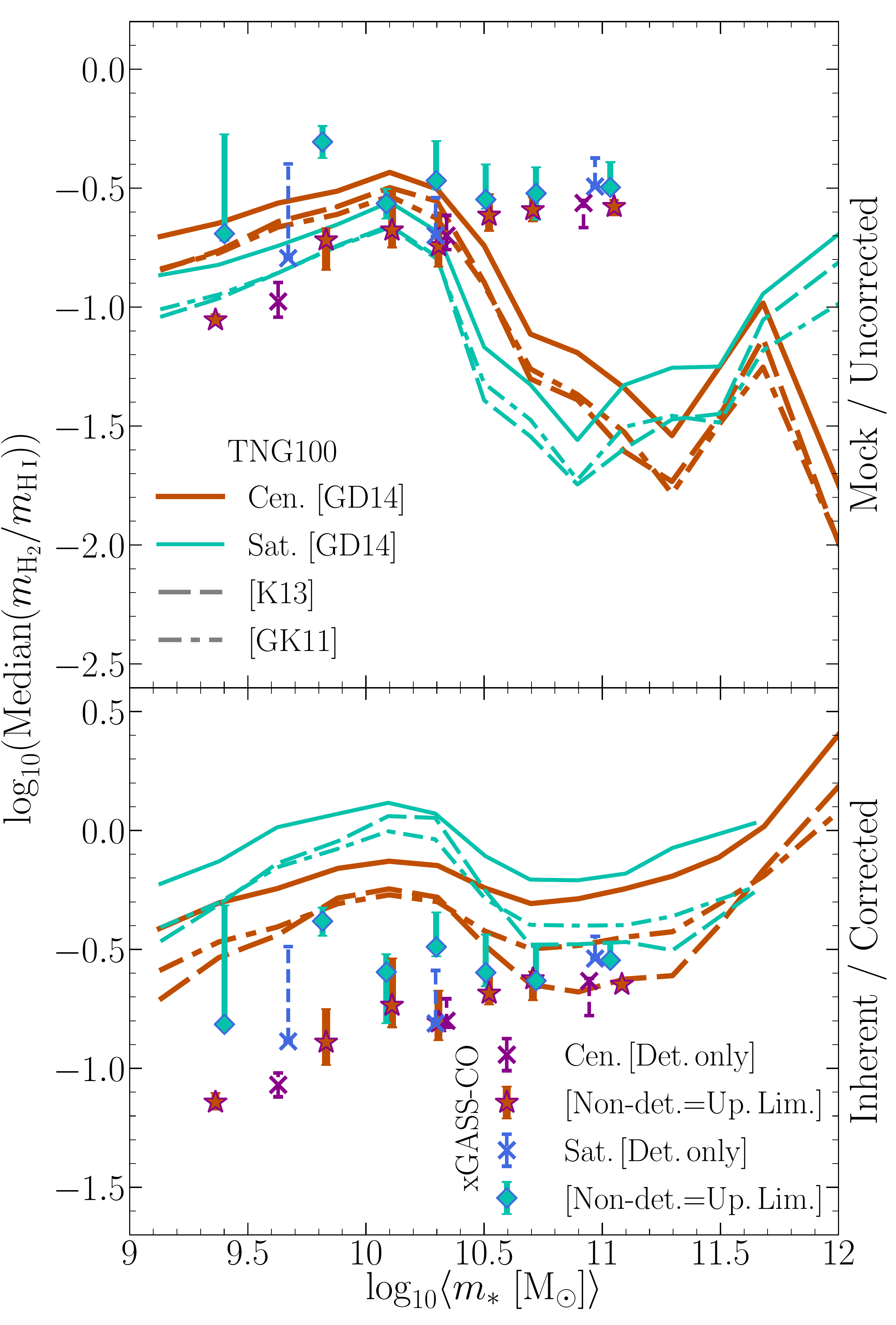}
\caption{Per Fig.~\ref{fig:H2HIFrac}, now exclusively showing median trends, with satellites and centrals separated.  Error bars on the xGASS-CO data are 68\% confidence intervals \emph{on the median} from bootstrapping.  The same stellar-mass bins are used for centrals and satellites, ensuring there are at least 15 galaxies of each type in each bin.  The low number of detected satellites in both xGASS and xCOLD GASS meant reducing the number of bins from seven with the inclusion of non-detections to three for the `detection only' data.}
\label{fig:H2HIFrac_SatCen}
\end{figure}

As has already been discussed here and in many other works, \Htwo~in satellite galaxies is expected to be stripped by ram pressure and/or tides systematically later after infall than \HI.  Based on this idea, one might expect that the \Htwo/\HI~mass ratios of satellites should be higher than centrals of the same stellar mass on average (although, this implicitly neglects other environmental and secular effects that might alter the \HI~and \Htwo~reservoirs of satellites differently).  We therefore compare the \Htwo/\HI~mass ratios of TNG100 centrals and satellites with xGASS-CO data in Fig.~\ref{fig:H2HIFrac_SatCen}.  
At first glance, the median \mHtwo/\mHI~points appear to be higher for satellites than centrals in xGASS-CO, albeit with potential overlap in some bins.  
To confirm this, we can follow the same procedure as we did for $m_{\rm H_2}/m_*$, i.e.~we can define a quantity $\Delta \log_{10} \! \left(m_{\rm H_2} / m_{\rm H\,{\LARGE{\textsc i}}} \right)$ that is analogous to Equation (\ref{eq:Delta}), swapping all instances of `$m_{\rm H_2}/m_*$' with `\mHtwo/\mHI'.  While we have omitted a figure analogous to Fig.~\ref{fig:Delta} for $\Delta \log_{10} \! \left(m_{\rm H_2} / m_{\rm H\,{\LARGE{\textsc i}}} \right)$ for brevity, we can again rule out the possibility that centrals and satellites have consistent distributions of $\Delta \log_{10} \! \left(m_{\rm H_2} / m_{\rm H\,{\LARGE{\textsc i}}} \right)$ through a KS test.  This holds regardless of whether we include non-detections at their upper limit or exclude them entirely.\footnote{The two-sided, two-sample KS statistic of $\Delta \log_{10} \! \left(m_{\rm H_2} / m_{\rm H\,{\LARGE{\textsc i}}} \right)$ for xGASS-CO centrals versus satellites when we include (exclude) non-detections in CO and \HI~is 0.31 (0.17), with a corresponding $p$-value of order $10^{-6}$ ($10^{-2}$).  These values are for the uncorrected \Htwo~masses, but including beam corrections does not lead to significant difference.}
The data therefore support the theory that satellites' \Htwo/\HI~mass ratios should be elevated relative to centrals.

Contrary to expectation, the mock-observed \Htwo/\HI~mass ratios of TNG100 satellites are \emph{lower} than centrals on average, as seen in the top panel of Fig.~\ref{fig:H2HIFrac_SatCen}.  This is, however, merely a reflection of the mocking procedure.  The applied beam for \HI~is typically greater in size than the extent of the simulated galaxies' \HI, while the opposite is true for \Htwo.  For satellites in particular, this artificially drives \mHI~up (see \citetalias{stevens19}), and thus drives the \mHtwo/\mHI~ratio down.  This is clarified by the bottom panel of Fig.~\ref{fig:H2HIFrac_SatCen}, where the inherent \mHtwo/\mHI~ratios of TNG100 galaxies are shown.  For fixed \HI/\Htwo~prescription, satellites in fact have systematically higher inherent \mHtwo/\mHI~ratios than centrals of the same mass.  This difference is notably greater at $m_*\!\lesssim\!10^{10}\,{\rm M}_\odot$ than at higher stellar masses, going from $\sim$0.2\,dex to effectively zero eventually.

% ============================================================ %

\subsection{Effect of parent halo mass on satellites}
\label{ssec:satenv}

Let us now assess the influence of environment on \Htwo~in a more quantitative sense.  Several metrics exist for quantifying galaxy environment. We use host halo mass as our metric, because (i) this is consistent with our previous work with TNG100 \citep{stevens19,stevens19b} and (ii) observational trends are evidently stronger with this metric than other popular alternatives, such as those that use the distance to the $N$th nearest neighbour \citep{brown17}.

\begin{figure}
\centering
\includegraphics[width=\textwidth]{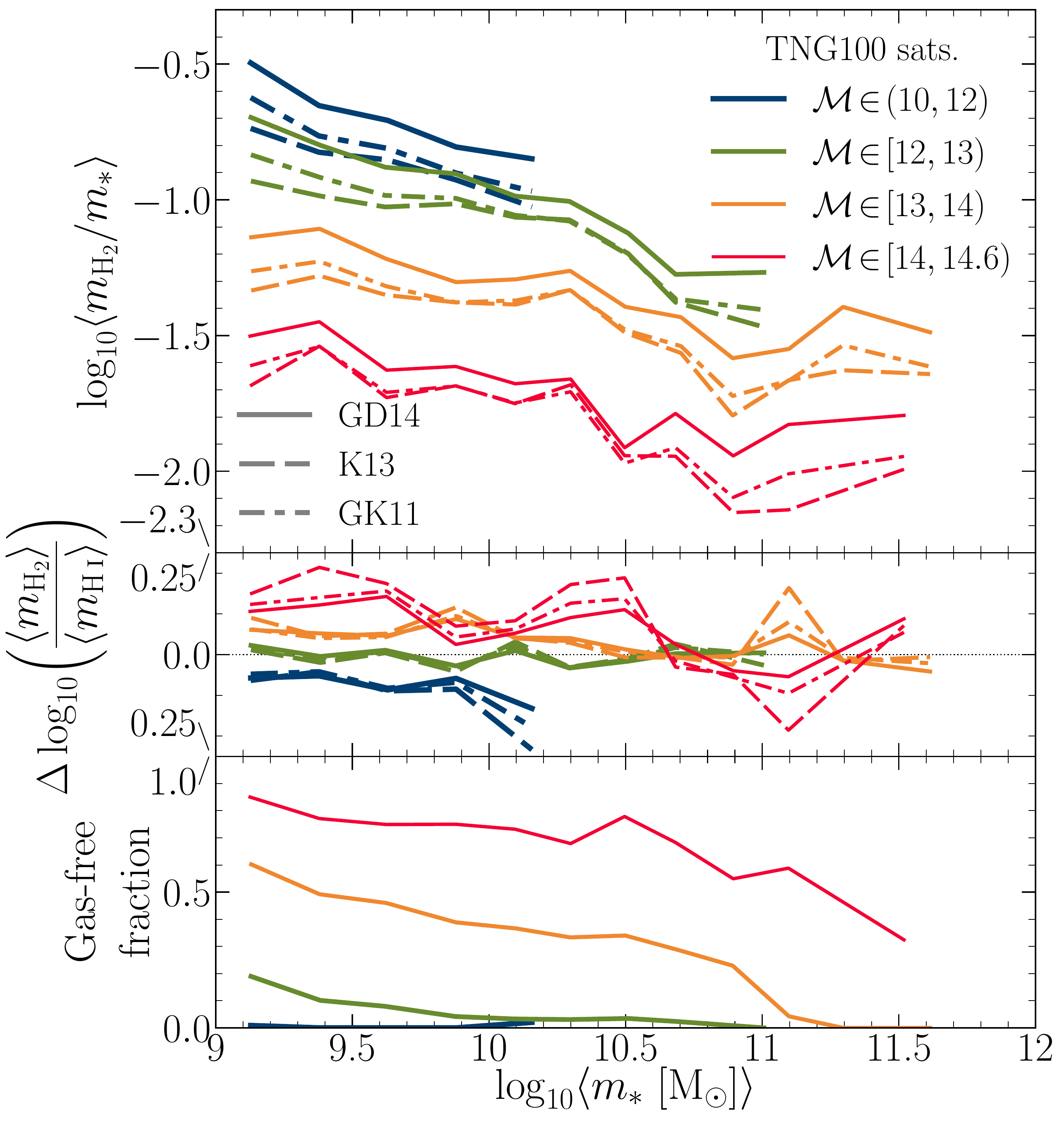}
\caption{
{\bf Top panel:} Mean inherent \Htwo~fraction of TNG100 satellites as a function of stellar mass for the same bins of parent halo mass used in \citetalias{stevens19}, where $\mathcal{M}\!\equiv\!\log_{10}\!\left(M_{\rm 200c}^{\rm parent}/{\rm M}_{\odot}\right)$.  Line colour and thickness relate to halo mass, while line style relates to the \HI/\Htwo~prescription.
{\bf Middle panel:} Ratio of the mean inherent \Htwo~mass to mean inherent \HI~mass for satellites in each given halo mass bin, \emph{normalised} by the same ratio for \emph{all} satellites at the same stellar mass (see Equation \ref{eq:DeltaMean}).  The $y$-scale has been stretched relative to the top panel for clarity.
{\bf Bottom panel:} Fraction of satellites for each halo mass bin that have no gas cells associated with them (and thus have $m_{\rm H\,{\LARGE{\textsc i}}} \! = \! m_{\rm H_2} \! = \! 0$).  The dependence of \Htwo~fraction on halo mass in the top panel is largely driven by this.  \HI/\Htwo~prescriptions are irrelevant for this panel.
}
\label{fig:H2Frac_Mean}
\end{figure}

In Fig.~\ref{fig:H2Frac_Mean}, we break TNG100 satellites into four bins of parent halo mass (in line with \citetalias{stevens19}), plotting the mean \Htwo~fraction as a function of stellar mass in the top panel.  For a given \HI/\Htwo~prescription, there is a clear decline in \mHtwo~in higher-mass haloes (denser environments) at fixed stellar mass; satellites hosted in haloes with $M_{\rm 200c}\!\geq\!10^{14}\,{\rm M}_{\odot}$ have a factor of $\sim$10 less \Htwo~than those in haloes with $M_{\rm 200c}\!<\!10^{12}\,{\rm M}_{\odot}$.
This result is very similar to that found for \HI~in \citetalias{stevens19} (see their fig.~6), implying that environmental effects in TNG100 have a clear and comparable signature on both atomic and molecular gas.  

Despite the signature of environment appearing similar for both \Htwo~and \HI~in TNG100,
we know from Section \ref{ssec:satcen} that the \Htwo/\HI~mass ratios of satellite galaxies in TNG100 are raised relative to centrals of the same stellar mass.
For this to be consistent, we should expect the average \Htwo/\HI~mass ratios of TNG100 satellites to have a residual dependence on parent halo mass as well.  Indeed, this dependence is present, as shown in the middle panel of Fig.~\ref{fig:H2Frac_Mean}.  The quantity on the $y$-axis of this panel has a somewhat subtle but deliberate definition.  For satellites in each halo- and stellar-mass bin, we measure the mean \Htwo~and mean \HI~mass, take the ratio of those means, log it, and from this subtract the same log ratio for all satellites at the same stellar mass.  That is,
\begin{multline}
\label{eq:DeltaMean}
\Delta \log_{10}\! \left( \frac{\langle m_{\rm H_2} \rangle}{\langle m_{\rm H\,{\LARGE{\textsc i}}} \rangle} \right) \equiv \log_{10}\! \left( \frac{\langle m_{\rm H_2}(m_*, {\rm sat.}, M_{\rm 200c}) \rangle}{\langle m_{\rm H\,{\LARGE{\textsc i}}}(m_*, {\rm sat.}, M_{\rm 200c}) \rangle} \right) \\
- \log_{10}\! \left( \frac{\langle m_{\rm H_2}(m_*, {\rm sat.}) \rangle}{\langle m_{\rm H\,{\LARGE{\textsc i}}}(m_*, {\rm sat.}) \rangle}  \right)\,.
\end{multline}
This definition removes systematic differences between the \HI/\Htwo~prescriptions (which can already be seen in the top panel) and allows for all galaxies with gas masses of zero to be naturally included [the latter would \emph{not} be the case were we to plot $\langle m_{\rm H_2}/m_{\rm H\,{\LARGE{\textsc i}}} \rangle$ or ${\rm Median}\!\left(m_{\rm H_2}/m_{\rm H\,{\LARGE{\textsc i}}}\right)$ instead, for example].
Evidently, parent halo mass only plays a noteworthy role in this quantity for $m_*\!<\!10^{10.5}\,{\rm M}_\odot$, and its role is rather noisy for $m_*\!\gtrsim\!10^{9.7}\,{\rm M}_\odot$.  The former resembles the fact that high-mass satellites are generally less prone to stripping, as they can often be comparable in size to their central.  Even for $m_*\!<\!10^{9.5}\,{\rm M}_\odot$, $\Delta \log_{10}\! \left( \langle m_{\rm H_2} \rangle / \langle m_{\rm H\,{\LARGE{\textsc i}}} \rangle \right)$ of satellites in the densest environments differs from those in the lowest-mass haloes by only $\sim$0.2\,dex.\\

What really drives the separation (or lack thereof) in the lines of different halo mass in the top two panels of Fig.~\ref{fig:H2Frac_Mean} is the fraction of galaxies that lack gas entirely.  This fraction is plotted in the bottom panel.  High-density environments in TNG100 are eventually able to fully remove the gas reservoirs of satellites.  An increase in the `gas-free' fraction has a clear signature in mean \Htwo~mass, while making it difficult for there to be significant differences between \HI~and \Htwo.
We should be clear that stripping is not actually binary in the simulation.  Rather, once the externally forced removal of gas reduces a galaxy to small number of cells, resolution effects take over, dooming the galaxy to lose its remaining few cells.  Up until this point, gas should be stripped smoothly.

Previous studies of galaxies' cold-gas content across environments have typically included much larger samples of observed galaxies than we have presented with xGASS-CO \citep[e.g.][]{stark16,brown17,sb17}; the important difference is they only had \HI~data on those galaxies.  With only 109 xGASS-CO satellites (including non-detections in CO), 
it is difficult to split the sample into bins of parent halo mass without losing statistical significance.  We therefore leave the results of Fig.~\ref{fig:H2Frac_Mean} as predictions for future observations to test.

% ============================================================ %

\subsection{Post-infall changes to satellites' gas}

\begin{figure}
\centering
\includegraphics[width=\textwidth]{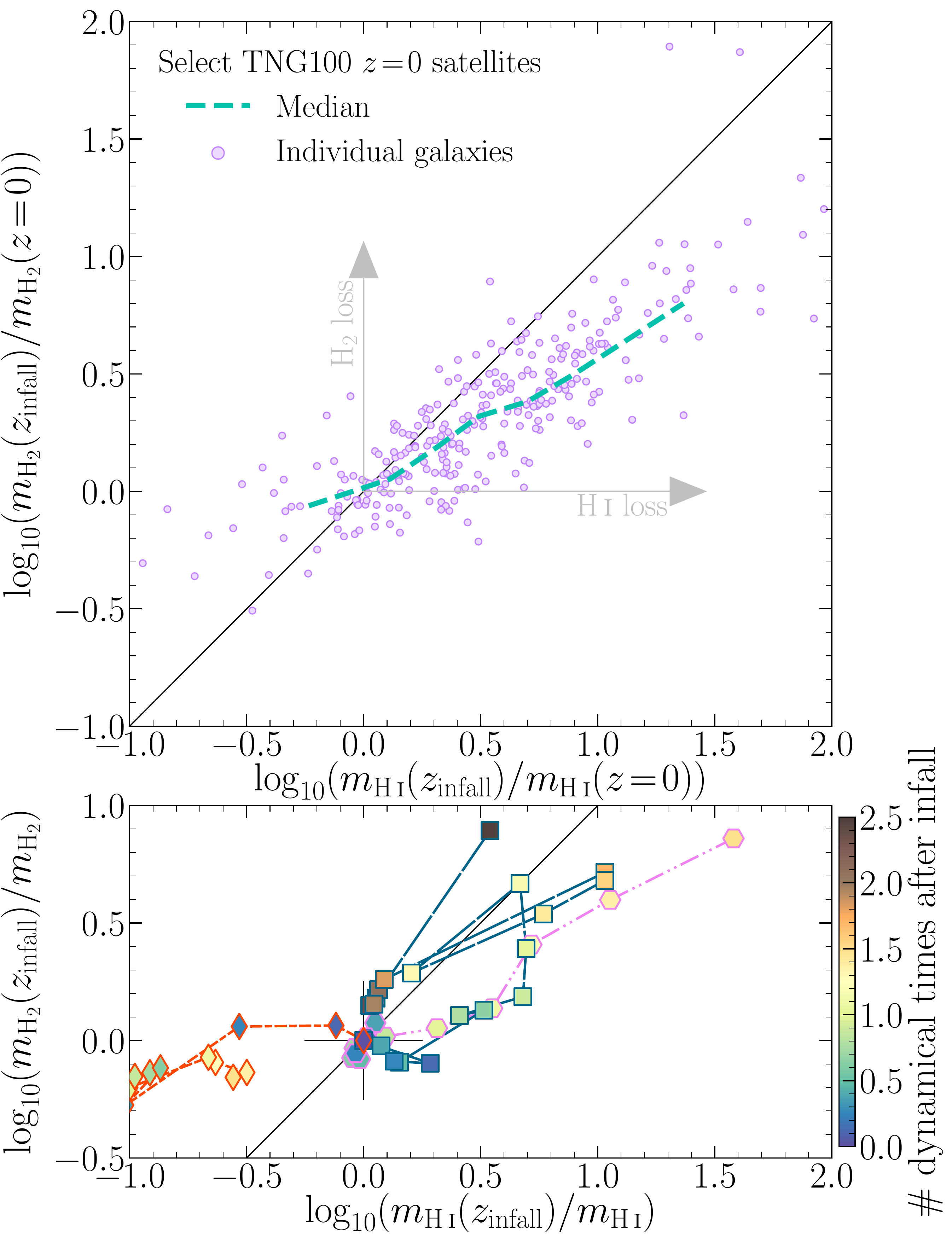}
\caption{Change in \HI~mass versus change in \Htwo~mass of TNG100 satellites since the first time they crossed $1.5\,R_{\rm 200c}$ of their parent halo (deemed the time of `infall').  
\HI~(\Htwo) losses after infall result in rightward (upward) movement in this space.
Satellites were selected at $z\!=\!0$ to have $m_*\!<\!10^{10.2}\,{\rm M}_{\odot}$, have an `\HI~deficiency' of 0.4\,dex (for their stellar half-mass radius), be within $1.5\,R_{\rm 200c}$ of their parent halo, and have been a {\sc subfind} satellite for no more than 3\,Gyr.  {\bf Top panel:} Points for individual galaxies and the running median are shown for the sample as it is at \zo.  {\bf Bottom panel:} Connected points of varied shape show tracks for three example galaxies after crossing $1.5\,R_{\rm 200c}$, coloured by time after infall.  The `dynamical time' for each satellite is calculated as $[10\,H(z_{\rm infall})]^{-1}$.  By definition, these tracks all start at (0,0) (the centre of the crosshair).}
\label{fig:HIH2_tlb}
\end{figure}

A primary utility of cosmological simulations is that galaxies can be tracked through time:~something that is observationally impossible for real galaxies.  With TNG100, we can directly see how the \HI~and \Htwo~reservoirs of galaxies change after infall (i.e.~once they have become satellites).
To this end, we select a sample of TNG100 satellite galaxies at \zo~with $m_*\!<\!10^{10.2}\,{\rm M}_{\odot}$ (avoiding AGN complications) that reside within $1.5\,R_{\rm 200c}$ of their parent halo, have been a satellite for $<\!3$\,Gyr, and have an \HI~mass that is between 0.4 and 2.0\,dex below the median for galaxies with the same stellar half-mass radius.  This last criterion means there is a nominal expectation that some stripping should have taken place (as the galaxies are `\HI~deficient'\,---\,e.g.~\citealt{boselli14c}), which we can test, while also ensuring they still have some gas left.  The requirement to have only been a satellite for 3\,Gyr minimizes any potential changes to the galaxies' gas properties associated with morphological transformation \citep[see the related discussions of][]{cortese19,joshi20}.  We track those galaxies back through the baryonic {\sc SubLink} merger trees \citep{rodriguez15} for TNG100, calculating their \HI~and \Htwo~content at each snapshot.\footnote{We note that not all snapshots have the fields saved that we require as input for our gas-phase decomposition methods.  We have therefore approximated the values for these missing properties, as described in Appendix \ref{app:app}, which we apply to \emph{all} snapshots for Fig.~\ref{fig:HIH2_tlb} (exclusively) to ensure we can fairly calculate \emph{changes} to the satellites' gas content over time.  The uncertainty introduced by these approximations is entirely negligible.}

In the top panel of Fig.~\ref{fig:HIH2_tlb}, we show the ratio of the sample satellites' \HI~mass at infall to that at redshift zero, versus the same ratio for \Htwo.  Compared in the bottom panel are three example evolutionary tracks of galaxies from their time of infall.  We define the `time of infall' here as that of the first snapshot at which the galaxy was both a satellite according to {\sc subfind} \emph{and} within $1.5\,R_{\rm 200c}$ of its host halo.  We chose $1.5\,R_{\rm 200c}$ rather than $1.0\,R_{\rm 200c}$ because environmental effects are not confined to $R_{\rm 200c}$; in fact, we could have selected an even larger radius, as environmental effects can extend out to $5\,R_{\rm 200c}$ in clusters \citep[see][]{bahe13,behroozi14,ayromlou19}.  Galaxies that have lost \HI~since infall will move to the right, while those that have lost \Htwo~will move upward.

Fig.~\ref{fig:HIH2_tlb} makes it clear that there is inhomogeneous evolution in satellites' \HI~and \Htwo~content after infall.  A relatively simple example of post-infall evolution is seen by the galaxy represented by hexagons in the bottom panel of Fig.~\ref{fig:HIH2_tlb}; this galaxy maintains its gas for several snapshots before gradually losing \HI~and \Htwo~at an increasing rate, moving towards the top-right of the panel.  Visual inspection of this galaxy (not shown here) clarifies that it is experiencing ram-pressure stripping.  The other two galaxies in the panel do not evolve so simply, reminding us that there are many physical processes that take place after infall, not just environmental effects.  The galaxy represented by the diamonds was in the process of experiencing a minor merger during infall.  This merger brought gas into the galaxy, causing both \mHI~and \mHtwo~to increase, shifting the galaxy to the bottom-left of the panel.  After this, its gas started to be stripped, causing a reversal in its parameter-space trajectory.  The galaxy represented by the squares began to be stripped after infall, but twice collided with intervening gas in the intrahalo medium (which likely would have otherwise been accreted by the central galaxy of the halo).  In the first instance, the galaxy was unable to hold the gas for longer than a single snapshot interval.  In the second instance, where the galaxy was closer to pericentre, it successfully accreted the gas it collided with, rejuvenating the galaxy for the better part of a dynamical time, before being stripped once again.

Despite the diversity of post-infall changes to satellites' gas, an underlying trend presents itself.  This is readily identified by the running median in the top panel of Fig.~\ref{fig:HIH2_tlb}, which is shallower than the 1:1 line, highlighting that losses in \HI~are generally more pronounced than in \Htwo.
In \citetalias{stevens19}, we argued that \Htwo~in TNG100 satellites might be stripped too efficiently, based on their SFRs being too strongly coupled to their \HI~content.  Here, we revise that interpretation somewhat; the preference for stripping \HI~appears to be present, but it may well be weaker than in the real Universe.

It is important to emphasize that \Htwo~need not be lost from satellites specifically by direct \emph{stripping}.  The three examples we described above were selected to be diverse rather than representative.  For other galaxies, it could feasibly be that \HI~is lost from ram pressure, but \Htwo~is consumed in star formation and simply unable to replenish \citep[e.g.][]{sb17}. However, the facts that (i) stripping manifests as the complete removal of satellites' gas in the densest environments in many cases (bottom panel of Fig.~\ref{fig:H2Frac_Mean}), (ii) depletion of \HI~in TNG100 satellites has a parallel effect on SFR \citepalias{stevens19}, and (iii) hydrodynamical forces in the simulation have no explicit knowledge of whether gas is predominantly atomic or molecular (as the phase decomposition is calculated in post-processing) imply that stripping affects all gas phases in the simulation.  Regardless of the physical mechanism(s) at play, Fig.~\ref{fig:HIH2_tlb} definitively shows us that \HI~is more readily depleted than \Htwo~(on average) once galaxies become satellites in TNG100.

% ============================================================ %

\subsection{Star formation as a proxy for molecular gas}
\label{sec:sf}

A wealth of empirical evidence points to a tight connection existing between the amount of molecular gas in a galaxy and its star formation rate, applying not only to integrated properties \citep[captured by a common \Htwo~depletion time\,---\,e.g.][]{saintonge17,tacconi18}, but also to localised surface densities \citep{bigiel08,leroy08}.  
After all, both molecule formation and star formation should occur in relatively dense, cold gas clouds. 
Observational measurements of galaxies' star formation rates are orders of magnitude more numerous than those of galaxies' \Htwo~content in the local Universe.
Comparing the star formation activity of TNG galaxies against those numerous data provides a less direct, but more statistically meaningful test of the simulated galaxies' molecular-gas content and its relation to galaxy environment.

Because the \HI/\Htwo~modelling was done in post-processing, the star formation law applied in TNG and the \Htwo~fractions we derive for the simulated galaxies do not have a precise one-to-one connection.  Unsurprisingly however, one does still find that TNG galaxies'  star formation rates and \Htwo~masses are closely correlated \citep{diemer19}, as is the case in other cosmological, hydrodynamic simulations \citep{lagos15b,rodriguez19,dave20}.  Comparison with xCOLD GASS shows that the SFR--\mHtwo~relationship in TNG is consistent with the real, local Universe \citep[fig.~4 of][]{diemer19}.
We have further found that there is no environmental dependence on this relationship in TNG100 (not shown here), which is important for the below.

To provide context for the results that follow, we present and discuss the $m_*$--sSFR plane of TNG100 galaxies in Appendix \ref{app:plane}.  We also refer the reader to the closely related work by \citet{donnari20a,donnari20b}, where the quenched fractions of TNG galaxies are investigated in detail, including environmental influences. 

\begin{figure}
\centering
\includegraphics[width=1.02\textwidth]{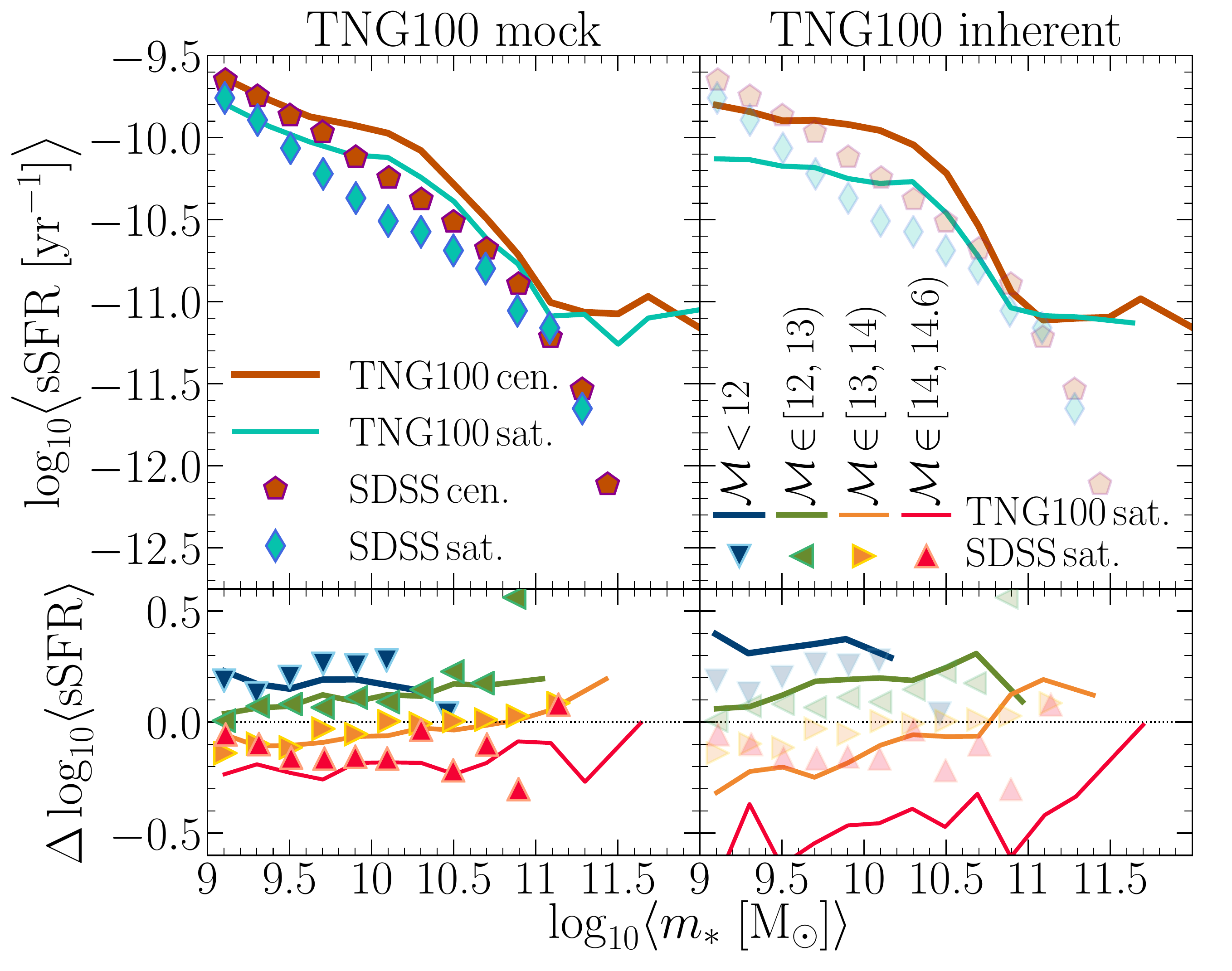}
\caption{Influence of environment on the stellar mass--specific star formation rate relation for TNG100 and SDSS galaxies.  The top panels show the running mean sSFR for centrals and satellites separately.
The left column measures TNG100 properties in a more appropriate manner for directly comparing to SDSS (as described in Sections \ref{ssec:other} and \ref{ssec:contam}).  The right-hand column shows the inherent results for TNG100.  Faded SDSS median points are shown for reference, but are not intended as a fair comparison in the right-hand panels.  The bottom panels break satellites into bins of parent halo mass, where the $y$-axis now shows the \emph{difference} in the mean sSFR of satellites in that halo mass bin from the total satellite population of the respective dataset (see Equation \ref{eq:Delta_sSFR}).  Note that the scale of the $y$-axis in this panel has been stretched for clarity.}
\label{fig:sSFR_satcen}
\end{figure}

In the two large panels of Fig.~\ref{fig:sSFR_satcen}, we show running means separately for satellites' and centrals' sSFRs as a function of stellar mass, for SDSS and both the mock (left panels, as described in Section \ref{ssec:other}) and inherent (right-hand panels) properties of TNG100.  
The left panels remain the definitive comparison with SDSS though.  
It is clear that satellites have systematically lower sSFRs than centrals of the same stellar mass (or, rather, a greater fraction of satellites are passive), both in the observed and simulated universes.  
Despite TNG100 generally having a smaller quenched population at most stellar masses (as can be inferred from Fig.~\ref{fig:sSFRplane})\,---\,which leads to differences in the shape of the  mean $m_*$--sSFR curves between the TNG100 mock predictions and SDSS\,---\,both datasets show similar \emph{relative} behaviour between centrals and satellites.  In particular, we can quantify the typical central--satellite separation in $\langle{\rm sSFR}\rangle$ across the full stellar-mass range as 0.19 and 0.15\,dex for SDSS and TNG100, respectively.

The key result of this subsection can be found in the bottom-left panel of Fig.~\ref{fig:sSFR_satcen}, where we split satellites by environment (quantified by their parent halo mass), and show the \emph{difference} between the mean sSFR of those satellites and that of \emph{all} satellites at the same stellar mass:
\begin{equation}
\label{eq:Delta_sSFR}
\Delta \log_{10}\!\left\langle{\rm sSFR}\right\rangle \equiv \log_{10}\!\left( \frac{\langle{\rm sSFR}(m_*, {\rm sat.}, M_{\rm 200c})\rangle}{\langle{\rm sSFR}(m_*, {\rm sat.})\rangle} \right)\,.
\end{equation}
At fixed stellar mass, the mean sSFR of satellites has a visible dependence on parent halo mass both in SDSS and TNG100 \citep*[for a directly related assessment of SDSS, see][]{wetzel12}.  If we are to trust that sSFR is a fair proxy for \Htwo~fraction, then Fig.~\ref{fig:sSFR_satcen} provides evidence\,---\,in addition to and independent from Fig.~\ref{fig:H2Frac_SatCen}\,---\,that environment-driven losses of \Htwo~in TNG are in quantitative agreement with the real Universe. 
Comparison of the two bottom panels of Fig.~\ref{fig:sSFR_satcen} highlights how, between satellites hosted in haloes with $M_{\rm 200c}\!<\!10^{12}\,{\rm M}_\odot$ and those with $M_{\rm 200c}\!\geq\!10^{14}\,{\rm M}_\odot$, a `true' difference in $\langle {\rm sSFR} \rangle$ of almost an order of magnitude can be reduced to a factor of $\sim$2.5 ($\sim$0.4\,dex) through the interpretation of and uncertainties associated with observational data. 
Comparison of the bottom-right panel of Fig.~\ref{fig:sSFR_satcen} with the top panel of Fig.~\ref{fig:H2Frac_Mean} also highlights that sSFR and \Htwo~are indeed very similarly affected by environment in TNG100.

% ============================================================ %
% ============================================================ %
% ============================================================ %
% ============================================================ %
% ============================================================ %
% ============================================================ %
% ============================================================ %

\section{Elephants swept under the rug}
\label{sec:elephants}
Before summarizing the findings of this paper, it is worthwhile touching on the underlying limitations at play, the likes of which do not always receive due attention.

\subsection{Resolution limitations}
Simulation outcomes are always at the mercy of resolution.  Of particular relevance here is the effect this can have on gas stripping in satellites, both from tides and ram pressure.  For example, a finite resolution limits the accuracy of hydrodynamical-force calculations between the interstellar medium of a satellite and the intrahalo medium it moves through.  While idealized simulations with {\sc arepo} suggest that the moving-mesh hydrodynamics scheme generally captures effects of ram pressure and induced Kelvin--Helmholtz instabilities more accurately than traditional SPH\footnote{Smoothed-Particle Hydrodynamics} codes \citep{springel10,hess12,sijacki12}, the median neutral gas cell number (neutral-gas mass divided by the baryonic mass resolution of $1.4\!\times\!10^6\,{\rm M}_\odot$) for TNG100 galaxies used in our study is only $\sim$1200.  The gas in many of the TNG100 galaxies is therefore more poorly resolved than the regimes of these tests.  The high fraction of gas-free satellites in TNG100 discussed in Section \ref{ssec:satenv} could potentially be influenced by this at some level.

Furthermore, the potential-well depth of galaxies can be affected by a host of numerical effects, including gravitational softening, the unequal mass partitioning of dark matter and baryons, the imposed density threshold for star formation, and the implementation of feedback \citep[see the recent results of][]{dutton19,ludlow19,ludlow20}.  This can artificially alter how susceptible a galaxy is to being stripped.
For example, we raised in Section \ref{ssec:mock} that the inner 2\,kpc of TNG100 galaxies are effectively `unresolved'.  Yet we showed a galaxy in Fig.~\ref{fig:images} with an \Htwo~half-mass radius of 2.2\,kpc, scarcely larger than the `resolution limit'.  
Fig.~\ref{fig:zdist} also shows that mock-observed \Htwo~masses can be sensitive to as little as the inner few kpc for a fraction of low-mass galaxies in TNG100.
These examples are by no means representative of all galaxies, but they anecdotally highlight that we are pushing TNG100 close to the limit of its capabilities.
In these situations, it also becomes especially difficult to capture environment-driven gas compression, where ram pressure might not only strip gas, but also lead to the conversion of \HI~into \Htwo~\citep[e.g.][]{henderson16}, which empirical evidence shows can happen \citep[e.g.][]{lee17}.

To elaborate on the non-trivial topic of resolution-based force limitations in the context of TNG would be worthy of a paper by itself, if not several.  
Where me may glean some insight regarding the impact of these limitations
in future is through applying a similar analysis in this work (and \citetalias{stevens19}) to the higher-resolution TNG50 simulation \citep{nelson19b,pillepich19}.

\subsection{Post-processing approximations}
Our post-processing models for performing the \HI/\Htwo~breakdown in TNG rely on common assumptions.  Namely, (i) they all assume some degree of equilibrium; (ii) they rely on the same principle ingredients:~local density, temperature, metallicity, and UV intensity; (iii) they are inherently two-dimensional models that have been translated to three dimensions using the imprecise Jeans approximation; and (iv) they are instantaneous in their application, meaning it is possible for discontinuities in the \Htwo/\HI~mass ratio of a given galaxy to occur from snapshot to snapshot.  The lattermost of these could artificially amplify the `random walk' nature of the tracks shown in the bottom panel of Fig.~\ref{fig:HIH2_tlb}.  Many of the limitations and assumptions of our \HI/\Htwo~modelling are discussed in greater detail in \citet{diemer18}.

\subsection{Observational uncertainties}
It should not be forgotten that observational data also carry potentially important limitations too.
Most notably, uncertainties in observed \Htwo~masses are dominated by systematics associated with the CO-to-\Htwo~conversion factor.
While the conversion factor adopted for xCOLD GASS is physically motivated and empirically calibrated \citep{accurso17}, it principally depends on metallicity, and secondarily depends on sSFR relative to the main sequence (which traces the UV field strength).  Ideally, the conversion would be done on local scales, but it is instead based on the galaxies' `integrated' properties, where `integrated' really means `inferred from the 3-arcsec SDSS fibre'.  The CO-to-\Htwo~conversion therefore implicitly assumes that the gas-phase metallicity and UV field are uniform in a given galaxy, which is almost certainly false.  As discussed in Section \ref{ssec:other}, the SDSS fibre size is far smaller than the beam size used to measure CO, and smaller still relative to the total extent of the xCOLD GASS galaxies.  Moreover, the diagnostics used to infer metallicity (oxygen abundance) from the SDSS emission line ratios carry significant systematic uncertainty themselves \citep*[see the review by][]{kewley19}.
Further systematics lie in the assumption that the gas-phase metallicity traces the dust-to-gas ratio (incidentally, this also applies to the \HI/\Htwo~prescriptions used for the simulation).

We note that we neglected to forward-model any \Htwo~mass uncertainties in our simulation predictions.  Had we done this, the scatter in \Htwo~fractions in the `mock' results would have increased to reflect the typical statistical error in the xCOLD GASS \mHtwo~measurements.  Given that systematics are even more important, it matters less that a predicted line goes through the data, and more that it is close and parallel.

In the spirit of forward-modelling, models that can be overlaid on simulations to infer CO emission provide a means to avoid converting CO flux from observations to \Htwo~mass entirely.  
While this requires thoughtful consideration of radiative transfer and chemistry to model in detail \citep[see e.g.][]{glover10},
even relatively straightforward methods to forward-model CO emission from cosmological-simulation galaxies
can lead to different conclusions than \Htwo-based comparisons with observations \citep{dave20}.
Some level of systematic uncertainty will likely still be present, but as with the mock-observation strategy employed in this paper, such a method is at least self-consistent and circumvents assumptions regarding structure and gradients in the observed galaxies.

% ============================================================ %
% ============================================================ %
% ============================================================ %
% ============================================================ %
% ============================================================ %
% ============================================================ %
% ============================================================ %

\section{In a nutshell}
\label{sec:conc}

We have studied the molecular-gas content of galaxies with $m_*\!\geq\!10^9\,{\rm M}_\odot$ at \zo~in the TNG100 simulation, focussing on how it depends on galaxy environment.  In addition to taking advantage of earlier work that developed and applied the post-processing framework to derive the \HI/\Htwo~properties of each gas cell in the simulation (\citealt{diemer18}; \citetalias{stevens19}), we have `mock observed' TNG100 galaxies to allow for the most direct comparison possible with data from the xCOLD GASS survey and SDSS.  Our main findings can be summarized as follows.

\begin{enumerate}

\item Were we to take the inherent properties of TNG100 and beam corrections of xCOLD GASS at face value, we would conclude that TNG100 galaxies are systematically too \Htwo-rich across all $m_*\!\geq\!10^9\,{\rm M}_\odot$ (by a factor of $\lesssim$2 at $m_*\!\lesssim\!10^{10}\,{\rm M}_\odot$, and greater at higher masses).
Most of the tension at low mass is relieved when comparing the mock-observed \Htwo~masses with the uncorrected xCOLD GASS masses instead, where the former includes the application of a Gaussian beam whose FWHM (consistent with the IRAM 30-m telescope) is smaller than the full extent of \Htwo~in the galaxies.  
As outlined in Section \ref{sec:comparison} and evidenced by Fig.~\ref{fig:corr}, the forward-modelling approach of this mock/uncorrected comparison with observations is more meaningful.
At $m_*\!>\!10^{10.5}\,{\rm M}_\odot$, the relative size of the mock beam is \emph{so} much smaller than the extent of molecular gas that inferred \Htwo~fractions in TNG100 drop by an order of magnitude. This result is summarized in Fig.~\ref{fig:H2Frac} with evidence for its reasoning seen in Fig.~\ref{fig:images}.  Similar statements can be made about the simulated galaxies' \mHtwo/\mHI~ratios (Fig.~\ref{fig:H2HIFrac}).

\item The onset of AGN feedback in TNG100 galaxies with $m_*\!\gtrsim\!10^{10.5}\,{\rm M}_\odot$ disrupts a significant fraction of gas discs, sometimes tearing them open from the inside out, displacing much of their neutral gas (e.g.~column 4 of Fig.~\ref{fig:images}).  This is more noticeable in satellite galaxies (Fig.~\ref{fig:H2Frac_SatCen}), which are less able to reaccrete their displaced gas than centrals, and are also more likely to have had merger-induced black-hole accretion in their histories by virtue of their being more clustered on average.  This feature was also identified in \HI~in \citetalias{stevens19}.
While the act of mock observation evidently washes this feature out, the finding that the \Htwo~half-mass radii of high-mass TNG100 galaxies are several times their stellar half-mass radii (when they should be comparable\,---\,cf.~\citealt{bolatto17,diemer19}; Fig.~\ref{fig:corr} of this paper) reaffirms that the TNG black-hole feedback model is too efficient at displacing gas as galaxies start to quench.

\item In TNG100, the median satellite is $\sim$0.6\,dex poorer in \Htwo~than the median central of the same stellar mass, per the simulated galaxies' inherent properties.  
However, in mock-observing the galaxies, we predicted that this signature would only be measured as $\sim$0.2\,dex in xCOLD GASS. 
The most significant aspect of the mocking that causes this smaller signature is the cross-contamination of centrals and satellites in the \citet{yang07} group catalogue.
In comparing the distribution of xCOLD GASS galaxies' \Htwo~fractions relative to the median of centrals at the same stellar mass, we showed that xCOLD GASS confirms our prediction (Fig.~\ref{fig:Delta}).

\item We found that the mean \Htwo~fraction of TNG100 satellites drops by an order of magnitude when comparing those in host haloes with $M_{\rm 200c} \! < \! 10^{12}\,{\rm M}_\odot$ to satellites in hosts with $M_{\rm 200c} \! > \! 10^{14}\,{\rm M}_\odot$ (Fig.~\ref{fig:H2Frac_Mean}).  While we could not compare this in a statistically significant way with xCOLD GASS, we used sSFR as a proxy for \Htwo~fraction to compare this environmental dependence with a sample of SDSS galaxies.  Again through tailored mock-observing, we predicted that SDSS should observe a $\sim$0.4\,dex difference in the mean sSFR of satellites hosted in $M_{\rm 200c} \! < \! 10^{12}\,{\rm M}_\odot$ haloes versus those hosted in $M_{\rm 200c} \! > \! 10^{14}\,{\rm M}_\odot$ haloes.  This is indeed in quantitative agreement with SDSS (Fig.~\ref{fig:sSFR_satcen}).

\item The canonical picture of environmental stripping of satellite galaxies' gas is that atomic gas (which is weighted towards the outskirts) is removed before molecular gas (which is deeper in the potential well).  
By tracking a sample of \zo~satellites back to their point of infall in Fig.~\ref{fig:HIH2_tlb}, we found it is indeed generally true for TNG100 galaxies that \HI~is lost at a higher fractional rate than \Htwo~after infall (although, plenty of secular effects continue to operate over this period as well).
This is also reflected by \zo~satellites having systematically elevated inherent \mHtwo/\mHI~ratios relative to centrals of the same stellar mass (Fig.~\ref{fig:H2HIFrac_SatCen}).  There is also evidence that the \mHtwo/\mHI~ratios of satellites at fixed stellar mass continue to increase as one looks at higher parent halo masses in TNG100.  But it is difficult to find a strong trend for this, as the fraction of satellites that host zero gas cells in the simulation also increases in denser environments; 75\% of satellites with $m_*\!\geq\!10^9\,{\rm M}_\odot$ hosted in haloes with $M_{\rm 200c}\!\geq\!10^{14}\,{\rm M}_{\odot}$ are devoid of gas entirely.  The cleanest trend we find is at $m_*\!\lesssim\!10^{10}\,{\rm M}_\odot$, where the ratio of the mean \Htwo~mass to mean \HI~mass of satellites is systematically $\sim$0.2\,dex higher in haloes with $M_{\rm 200c}\!\geq\!10^{14}\,{\rm M}_{\odot}$ than those with $M_{\rm 200c}\!<\!10^{12}\,{\rm M}_{\odot}$ (Fig.~\ref{fig:H2Frac_Mean}).  This is a prediction that could be tested with stacking of CO and \HI~spectra targeted at the right galaxies.

\end{enumerate}

To study environmental effects on galaxies' \Htwo~in greater detail, a large-scale, blind CO survey(s) is ultimately needed.
While the near future is bright for \HI~surveys of this nature, which will lead to a wealth of environment studies (among other topics\,---\,e.g. \citealt{adams19,koribalski20,maddox20}), equivalents for CO will inevitably lag behind.  
In the interim, to further advance observational tests of the environmental dependence of galaxies' \Htwo~from theoretical models and simulations, we might be better off targeting a handful of galaxy clusters, and studying their members in finer detail.
Recently completed surveys like VERTICO (Brown et al.~in prep.) will prove useful in this regard, offering a three-dimensional view of 51 Virgo cluster galaxies in CO emission.

% ============================================================ %
% ============================================================ %
% ============================================================ %
% ============================================================ %
% ============================================================ %
% ============================================================ %
% ============================================================ %

\section*{Acknowledgements}
All plots in this paper were built with the {\sc matplotlib} package for {\sc python} \citep{hunter07}.  Our analysis also made extended use of the {\sc numpy} package \citep{numpy}.

ARHS acknowledges receipt of the Jim Buckee Fellowship at ICRAR-UWA.
Parts of this research were supported by the Australian Research Council Centre of Excellence for All Sky Astrophysics in 3 Dimensions (ASTRO 3D), through project number CE170100013.
LC is the recipient of an Australian Research Council Future Fellowship (FT180100066) funded by the Australian Government.
FM acknowledges support through the program `Rita Levi Montalcini' of the Italian MIUR.

\section*{Data availability}
The IllustrisTNG simulations are publicly available at \url{https://www.tng-project.org/data/}, with a full description of those data provided by \citet{nelson19}.  The \HI/\Htwo~properties in the public database are not identical to those used in this paper, but they are based on the same methodology and are known to agree well (see section 2.3 of \citetalias{stevens19}).  Integrated galaxy properties used in this paper are also not directly provided in the public database, but are recoverable through the particle data in principle.  Otherwise, data used in this paper can be made available upon request.

The xCOLD GASS survey data are publicly available at \url{http://www.star.ucl.ac.uk/xCOLDGASS/data.html}.  Similarly, data for xGASS are available at \url{https://xgass.icrar.org/data.html}.

SDSS data from the MPA--JHU catalogue are available at \url{https://wwwmpa.mpa-garching.mpg.de/SDSS/DR7/}, with improved stellar masses founds at \url{http://home.strw.leidenuniv.nl/\~jarle/SDSS/}.

% ============================================================ %
% ============================================================ %
% ============================================================ %
% ============================================================ %
% ============================================================ %
% ============================================================ %
% ============================================================ %

\appendix

\section{Ancillary results}
\label{app:anc}

\subsection{The stellar mass--specific star formation rate plane}
\label{app:plane}

\begin{figure}
\centering
\includegraphics[width=1.435\textwidth,angle=270]{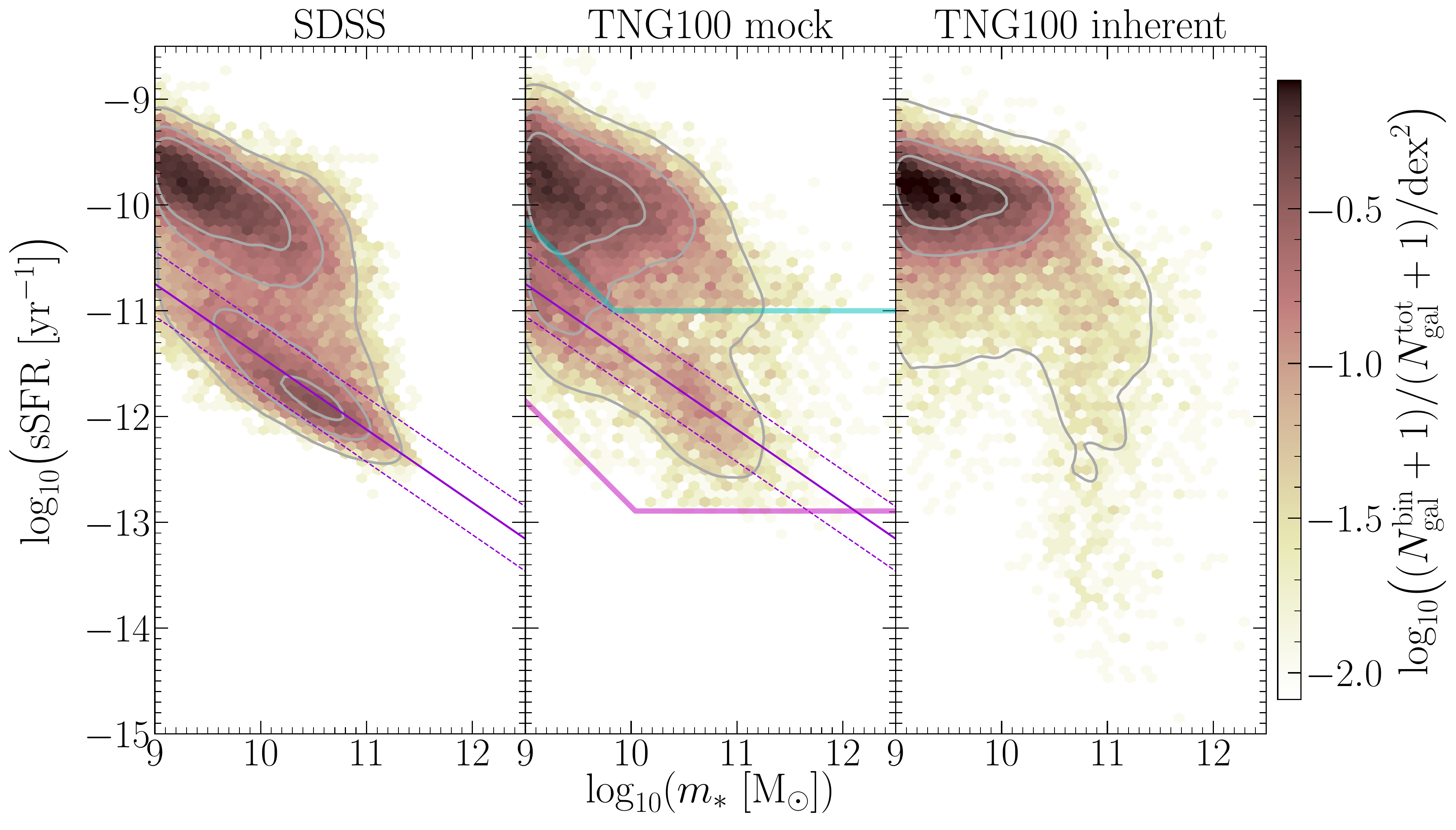}
\caption{Stellar mass--specific star formation rate plane for SDSS and TNG100 galaxies.  Pixel colour indicates the normalised number density of galaxies.  Grey contours encompass 38, 68, and 95\% of the galaxies contained in the axes (approximately equivalent to 0.5$\sigma$, 1$\sigma$, and 2$\sigma$ contours, respectively, based on a two-dimensional Gaussian kernel density estimator applied to the underlying galaxy sample).  
{\bf The top panel} shows a volume-limited subsample of SDSS galaxies with $m_*/{\rm M}_{\odot} \!\in\!\left[10^9,10^{11.5}\right]$ in the redshift interval $[0.02,0.05]$.  The violet, solid line shows a fit to the quiescent population (slope, intercept = $-0.689$, $-4.547$).  The dashed, violet lines show the vertical standard deviation of the population about that fit ($\sigma\!=\!0.305$).  
{\bf The bottom panel} shows TNG100 galaxies, per their inherent properties, where SFR is summed from the gas cells' instantaneous values.  A significant population of galaxies with ${\rm SFR}\!=\!0$ is not seen in this panel (nor do the contours account for them).  
{\bf In the middle panel}, TNG100 galaxy properties are remeasured to be more directly comparable with SDSS.  First, star-forming galaxies (instantaneous ${\rm sSFR} \! \equiv \! {\rm SFR}/m_* \! > \! 10^{-11}\,{\rm yr}^{-1}$) have their SFR remeasured on a 20-Myr historical time-scale, based on the birth masses of the galaxies' star particles that formed within that look-back period.  Any galaxy with either an instantaneous or 20-Myr ${\rm SFR}\!<\! 10^{-11}\,{\rm yr}^{-1}$ has its SFR remeasured on a 1-Gyr time-scale.  The baryonic mass resolution of TNG100 (taken as $1.4\!\times\!10^6\,{\rm M}_{\odot}$) means any 1-Gyr sSFR must lie rightward of (or close to) the thick, transparent, magenta line.  Similarly, any galaxy with a measured 20-Myr SFR must lie rightward of the cyan line.  Finally, any remaining galaxies with ${\rm SFR}\!=\!0$ have a random sSFR pulled from a Gaussian distribution that depends on their stellar mass and matches the fitted SDSS quiescent population; the violet lines are shown again in this panel for reference.}
\label{fig:sSFRplane}
\end{figure}

Similar to the purpose of Section \ref{sec:comparison} for \Htwo, to provide context for the effect of environment on SFRs in TNG galaxies, it is informative to briefly review the overall star formation activity of TNG galaxies as compared to galaxy survey data.  We thus plot the stellar mass--specific star formation rate plane for TNG100 alongside SDSS in Fig.~\ref{fig:sSFRplane}.  While this plane and the related colour--mass plane have been presented for the TNG suite in several papers already \citep{nelson18,weinberger18,donnari19,pillepich19,zanisi20,zinger20}, our inclusion of Fig.~\ref{fig:sSFRplane} not only satisfies self-containment, but it is also value-added through its inclusion of SDSS data and a purpose-made `mock' plane for TNG100, as described in Section \ref{ssec:other}.
The sample of SDSS galaxies is described in Section \ref{ssec:sdss}.

The presence of a bimodality in the SDSS $m_*$--(s)SFR plane is well documented \citep[e.g.][]{brinchmann04,chang15} and is clearly visible in the top panel of Fig.~\ref{fig:sSFRplane}.  A direct comparison of TNG100 (based on instantaneous SFRs) in the bottom panel suggests the simulation is almost missing a bimodality entirely.  This is, however, merely a reflection of the fact that 20.5\% of TNG100 galaxies with $m_*\!\geq\!10^9\,{\rm M}_{\odot}$ have an instantaneous ${\rm SFR} \! = \! 0$.  This contrasts to SDSS, where $<$1.5\% of the sample we use do not have a non-zero, measured SFR.  We also need to be conscious of the time-scales on which SFRs are measured in SDSS.  We aim to match this in the middle panel of Fig.~\ref{fig:sSFRplane} by using mock SFRs (and stellar masses) as described in Section \ref{ssec:other}.
With this, it is visually much clearer that there is a distinct quenched population of TNG100 galaxies (for a discussion surrounding the \emph{colour} bimodality in TNG\,---\,which falls out without artificial assignment as done here\,---\,see \citealt{nelson18}).  
But, loosely speaking, based on the difference in contours between the top and middle panels of Fig.~\ref{fig:sSFRplane}, there are apparently still too few quenched galaxies at most stellar masses in TNG100 relative to SDSS.  For an elaborate discussion on this topic, see \citet{donnari20a}.

% ============================================================ %

\subsection{\HI~and neutral fractions of galaxies}
\label{app:HI}

\begin{figure}
\centering
\includegraphics[width=0.9455\textwidth]{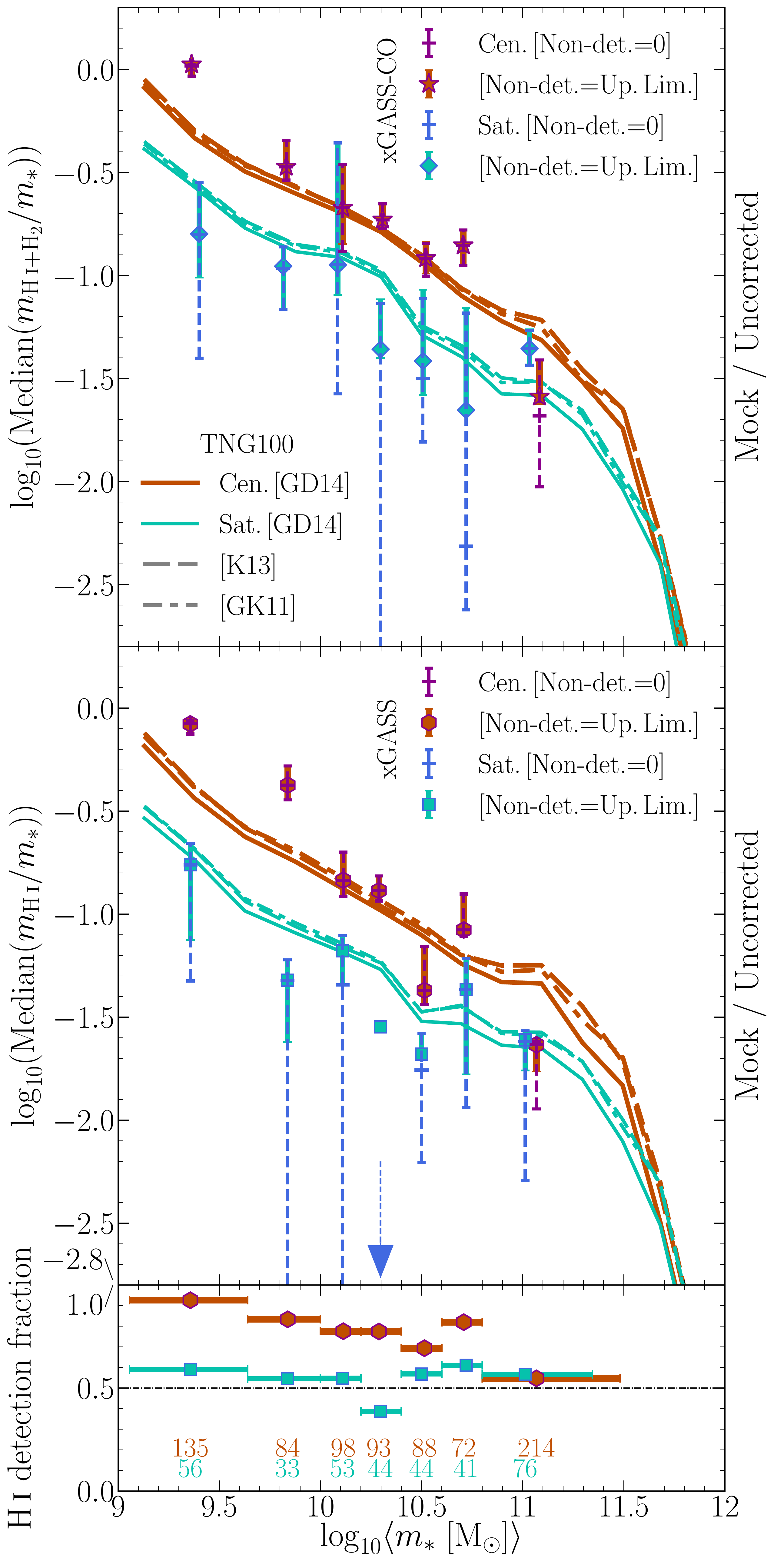}
\caption{{\bf Top panel:} Median neutral-hydrogen fraction as a function of stellar mass for TNG100 (at \zo) and xGASS-CO centrals and satellites.
{\bf Middle panel:} Median atomic-hydrogen fraction as a function of stellar mass, where now TNG100 is compared against the full xGASS sample.  The same bins and plotting style are used as in Fig.~\ref{fig:H2Frac_SatCen}.  Error bars on observational data are uncertainties \emph{on the median} from bootstrapping.  All TNG100 results have been updated from \citetalias{stevens19} to adhere to the mock-observing method in Section \ref{ssec:mock} of this paper (cf.~figs 4 \& 5 from \citetalias{stevens19}).
{\bf Bottom panel:} Detection fraction of the 21-cm line in each xGASS bin.}
\label{fig:HI}
\end{figure}

Because several aspects of the mock procedure outlined in Section \ref{ssec:mock} were updated from \citetalias{stevens19}, here we reproduce the key comparisons between TNG100 and xGASS from \citetalias{stevens19}, folding in these new aspects.  Fig.~\ref{fig:HI} is analogous to the left panels of Fig.~\ref{fig:H2Frac_SatCen}, and shows the running medians for galaxies' neutral-hydrogen and atomic-hydrogen fractions.  Because the mock-observed atomic-hydrogen fractions of TNG100 are compared against the full xGASS sample, we retain the mock redshifts used in \citetalias{stevens19} for this panel (see section 3.1 and the upper panels of fig.~2 of \citetalias{stevens19}).  No beam corrections are used for the observations in either panel.

Despite the larger sample size of xGASS, significant uncertainties remain on the satellites medians, as the detection fraction for \HI~hovers at 50\% for satellites of all masses.  This is shown in the bottom panel of Fig.~\ref{fig:HI}.  Nevertheless, the systematic distinction between the \HI~content of centrals and satellites is much clearer than was the case for \Htwo.

% ============================================================ %
% ============================================================ %
% ============================================================ %
% ============================================================ %
% ============================================================ %
% ============================================================ %
% ============================================================ %

\section{Approximate \HI/\Htwo~calculation for `mini snapshots'}
\label{app:app}

Nominally, the neutral fraction of all non-star-forming (NSF) gas cells in TNG comes from the recorded values in the snapshot outputs.  However, for the majority of snapshots, only a subset of the particle/cell fields were saved for the sake of reducing the simulation's storage requirements.  Neutral fraction is one of the fields lost in these so-called `mini snapshots'.  To circumvent this, we find a polynomial fitting function for the neutral-fraction field at \zo~in terms of the `electron abundance' field, $f_e$, which \emph{is} saved for all snapshots:
\begin{subequations}
    \label{eq:fn_NSF}
    \begin{equation}
        f_n^{\rm NSF} \simeq 
        \left\{
        \begin{array}{lr}
            \xi_1 + \xi_2\, f_e & \forall f_e \leq f_{e0}\\
            \xi_3 + \xi_4\, f_e + \xi_5\, f_e^2  & \forall f_e > f_{e0}
        \end{array}
        \right.
        \,,
    \end{equation}
    \begin{equation}
        \xi_3 = \xi_1 + (\xi_2- \xi_4)\,f_{e0} - \xi_5\,f_{e0}^2\,,
    \end{equation}
\end{subequations}
where $f_n$ is neutral fraction, $f_e$ is formally the ratio of free electrons to hydrogen protons (including free protons + those in hydrogen atoms and molecules, but excluding those in any other chemical species).  The best-fitting values for the five parameters are given in Table \ref{tab:lp1}.  These parameters have a bimodal dependence on the specific-energy range of the particles.  For $u \! > \! 10^{8.7}\,{\rm m\,s}^{-1}$, almost all hydrogen (and other species) is ionized, leaving $f_e \! \gtrsim 1$.  Below this energy limit, $f_e$ spans the full possible range.  

\begin{table}
    \centering
    \begin{tabular}{l | r r r r r}\hline
        $u$ & $\xi_1$ & $\xi_2$ & $\xi_4$ & $\xi_5$ & $f_{e0}$\\\hline
        $< \! 10^{8.7}\,{\rm m}^2\,{\rm s}^{-2}$ & 0.995 & -0.907 & -4.562 & 1.854 & 0.986\\
        $\geq \! 10^{8.7}\,{\rm m}^2\,{\rm s}^{-2}$ & 0.998 & -0.996 & -1.943 & 0.859 & 0.984\\\hline
    \end{tabular}
    \caption{Best-fitting values for Equation (\ref{eq:fn_NSF}), rounded to three decimal places.  The first column gives the range of internal energies per unit mass for the cells.}
    \label{tab:lp1}
\end{table}

The abundances of individual chemical species are also not recorded in the mini snapshots.  This includes the hydrogen abundance, $X$.  But the total metallicity, $Z$, is available.  We therefore again find an approximate fitting function for $X(Z)$ at \zo:
\begin{equation}
X \simeq 
\left\{
\begin{array}{lr}
0.753 - 1.26\, Z & \forall Z \geq 0.025\\
0.762 - 2.30\, Z + 24.2\, Z^2  & \forall Z < 0.025
\end{array}
\right.
\,.
\label{eq:XZ}
\end{equation}
We ensure an upper limit of $X\!\leq\!0.76$ always, as this was the initial value assumed in the simulation.

Our approximations for $f_n$ and $Z$ both carry an absolute error of less than 0.001 when considering all cells in all haloes used in this work.  For gas cells with large neutral masses, Equation (\ref{eq:fn_NSF}) tends to over-predict $f_n$ by 0.005.  Even then, the uncertainty introduced by these approximations is definitively negligible.
With these approximations, we can otherwise follow the same method for calculating \mHI~and \mHtwo~for galaxies per \citetalias{stevens19}.  Fig.~\ref{fig:HIH2_tlb} is the only place in this paper where we have to use this approximation method.  {\sc python} routines for calculating neutral fractions and metallicity in line with this method are publicly available on Github.\footnote{See the {\tt neutralFraction\_from\_electronFraction()} and {\tt HI\_H2\_masses()} functions at \url{https://github.com/arhstevens/Dirty-AstroPy/blob/master/galprops/galcalc.py}.}

\label{lastpage}
\end{document}